\documentclass[prd,superscriptaddress,nofootinbib,showpacs,preprint]{revtex4}

\usepackage{graphicx} 
\usepackage{graphics}
\usepackage{amsmath,amssymb} 
\usepackage{array}
\usepackage{epsfig}
\usepackage{epstopdf}
\usepackage[colorlinks]{hyperref}  
\usepackage{hyperref} 
\usepackage[usenames,dvipsnames]{color} 
\usepackage{ulem}
\begin{document}
\title{Threshold effects and renormalization group evolution of neutrino
parameters in  TeV scale seesaw models \\}
\author{Srubabati Goswami}
\email{sruba@prl.res.in}
\affiliation{Physical Research Laboratory, Ahmedabad-380009, India}
\author{Subrata Khan}
\email{subrata@prl.res.in}
\affiliation{Physical Research Laboratory, Ahmedabad-380009, India}
\author{Sasmita Mishra}
\email{sasmita@prl.res.in}
\affiliation{Physical Research Laboratory, Ahmedabad-380009, India}

\begin{abstract}
We consider the threshold effect on the renormalization 
group (RG) evolution of the neutrino masses and mixing angles 
in TeV scale seesaw models. We obtain the analytic expressions 
using the factorization method in presence of threshold effects. 
We also perform numerical study of RG effects in two specific 
low scale seesaw models following the bottom-up approach and 
ascertain the role of seesaw thresholds in altering the values of 
masses and mixing angles during RG evolution. 
 
%The models considered by  us are such that it is possible to 
%reconstruct the Yukawa coupling matrices from the low energy
%neutrino oscillation data. 

\end{abstract}

\maketitle

\section{Introduction}
\label{sec:Intro}
Neutrino oscillation experiments 
have established that this elusive particle
has tiny but nonzero mass and different flavor states mix 
amongst each other.
The oscillation experiments can only 
measure the mass squared differences. For three neutrino flavors
these are measured as:   
$\Delta m_{21}^2 \simeq10^{-5}\,{\rm eV}^2$,
and $|\Delta m_{31}^2|\simeq 10^{-3}\,{\rm eV}^2$,
where $\Delta m_{ij}^2 = m_i^2 -m_j^2$, and $m_i$ are the mass eigenvalues.
On the other hand, cosmological observations 
provide an upper bound on 
the sum of masses of the neutrinos as: 
$\sum m_i \le 0.23 \, {\rm eV}$ \cite{Ade:2013ktc}. 

%For massive particles, mass eigenstates are in general  
%linear combinations of flavor eigen states, giving rise to the   
%mixing matrix, $U_{\rm PMNS}$, where PMNS stands for
%Pontecorvo-Maki-Nakagawa-Sakata. For three neutrino
%flavors this is parametrized by three mixing angles 
%$\theta_{ij}$ and three phases. 
%It is
%known from experiments that two of the three mixing angle are large:
%$\sin^2 \theta_{12} \simeq 0.32,\,  \sin^2 \theta_{23}\simeq 0.6$  
%whereas the third one is smaller; $\sin^2 \theta_{13}\simeq 0.025$. 
%The current best fit values are given in 
%Table(\ref{tab:oscptm}) \cite{Tortola:2012te}. 
%The phases still remain unknown.

 The smallness of the 
neutrino mass can be explained by type-1 seesaw mechanism  
 \cite{Minkowski:1977sc,Yanagida:1979as,GellMann:1980vs,Glashow:1979nm, Mohapatra:1979ia}
in which one adds heavy right handed neutrinos to the Standard 
Model (SM) for its ultraviolet completion.  
In canonical type-I seesaw model,
the heavy neutrino mass scale  is at 
$\mathcal{O}\left(10^{14}\,\text{GeV}\right)$ in order to 
generate the small masses of the light neutrinos. 
Since the Large Hadron Collider (LHC) has started running the 
subject of TeV scale seesaw which can produce observable signature 
at the LHC have invoked considerable interest. 
In the context of standard type-I seesaw, lowering the seesaw scale 
is only possible for certain specific textures 
\cite{Adhikari:2010yt,Kersten:2007vk, Pilaftsis:1991ug}. 
A more natural way is provided by the  mechanisms of inverse 
\cite{Mohapatra:1986bd}
or linear \cite{Gu:2010xc, Zhang:2009ac,Hirsch:2009mx} seesaw in 
which one adds additional singlets to the theory.  In these models
small neutrino mass is generated by smallness of the lepton number 
violating coupling of the singlets and the seesaw scale 
can be at TeV  in general.

Within the context of seesaw mechanism the neutrino masses and mixing 
angles arise from the dimension 5 effective operator term 
$\kappa l l \phi\phi$, where $l$'s are the lepton doublets,
$\phi$ is the SM Higgs field, $\kappa \sim a/M$. $a$ being a dimensionless parameter and  
$M$ is the mass scale where the
additional fields are integrated out \cite{Weinberg:1979sa}.
This generates a neutrino mass, $m_\nu \sim \kappa v^2$
once the Higgs field gets Vacuum Expectation Value (VEV), $v$. 
In general, $\kappa$ is a matrix and upon diagonalization
it gives the mass eigenvalues.

%Whereas the information about
%mixing angles and phases are contained in the matrix, $U_{\rm PMNS}$. 
%For neutrino mass $0.1$ eV and for a typical Higgs vacuum expectation
%value $100$ GeV the mass scale of the extra fields is $M\sim 10^{14}$ GeV.
%It is interesting to see that the operator violates lepton number
%by two units. So the large suppression of neutrino mass via heavy fields 
%gets related to their Majorana nature. This feature elegantly relates
%the seesaw mechanism with baryogenesis via leptogenesis mechanism in the
%Universe. Unfortunately both the features; smallness of neutrino mass
%and viability of leptogenesis require a heavy Majorana scale which
%make it impossible for direct verification of seesaw mechanism
%via production of heavy neutrinos.  

Note that the current neutrino oscillation parameters are 
measured at low energy but the dimension 5 operator emerges at the seesaw
scale which is usually high. Thus this is  governed
by the effects of renormalization group (RG) evolution between
the seesaw scale and the low energy scale \cite{Babu:1993qv,Antusch:2001ck}. 
This induces radiative corrections to the  
the leptonic mixing angles, phases
and masses. 
%The effect of
%RG running below the seesaw scale is different from that above the 
%seesaw scale.
Below the seesaw scale the RG evolution of the 
neutrino mass operator is governed by the effective theory which
is same for all seesaw models. But above the seesaw scale one
has to consider the full theory.
This region can be
model dependent.  

In this paper  we have studied
the dynamics of neutrino parameters in  low scale seesaw models
such that at least one of the heavy particles is present in the 
theory till the scale is about 1 TeV. This would then introduce 
threshold corrections to the RG evolution of the neutrino masses
and mixing. We obtain analytic expressions for radiative 
corrections to  neutrino masses and 
mixing angles in presence of seesaw threshold effects. 
%These are obtained for the simplified case of tri-bimaximal 
%mixing at the high scale which requires $\theta_{13}=0$.
Analytic expressions for modified neutrino masses and mixing 
angles induced by RG running have also 
been obtained in \cite{Bergstrom:2010id,Bergstrom:2010qb} 
for single and multiple 
thresholds. We use the factorization 
method outlined in \cite{Ellis:1999my, Chankowski:2001mx} 
and applied in connection to neutrino masses and mixing 
angles to obtain the RG corrected expressions in \cite{Dighe:2007ksa,Dighe:2006sr,Dighe:2006zk}. 
This method 
allows one to relate the low and high scale parameters 
without actually solving the beta functions. 
%The analytic expressions 
%for the running of neutrino masses and mixing angles obtained  
%in \cite{Dighe:2007ksa,Dighe:2006sr,Dighe:2006zk,Goswami:2009yy}, 
%were, without the inclusion of seesaw threshold effects. 
%We obtain the modifications to these expressions  
%after including threshold effects.
To gauge the impact of the 
threshold effects we perform a numerical study incorporating these.
%One of our aims is to see if the incorporation of threshold effects
%can cause any significant running of $\theta_{13}$. 
%We also do numerical studies of   
%the effect of RG evolution 
%including threshold effects. 
Such a study requires the knowledge of the neutrino Yukawa matrix ($Y_\nu$). 
We adopt bottom-up running approach and consider two models 
for which it is easy to reconstruct the neutrino Yukawa matrices from the 
low scale values of neutrino masses and mixing angles. 
The first case is the  linear seesaw model studied in \cite{Gavela:2009cd}. 
For this, the minimal model needs two right handed singlets  which are 
pseudo-Dirac in nature.
This results in a hierarchical
spectrum of light neutrino masses with one of the mass eigenvalues 
as zero. It is well known that RG effects are small for hierarchical 
neutrinos. We aim to find out how much this conclusion is altered in 
a seesaw model where the heavy state is connected to the theory up to
the TeV scale. 
The second case that we consider, constitutes of a quasidegenerate
mass spectrum of light neutrinos  that can come from addition of
three right handed heavy
neutrinos degenerate in masses. Here we construct the Yukawa
coupling using the Casas-Ibarra parameterization \cite{Casas:2001sr}.

%.The RG running of neutrino mass
%matrix also requires the running of associated Yukawa couplings, gauge couplings
%and Higgs self coupling. Since the form of neutrino Yukawa coupling is unknown 
%in this case, we reconstruct the same from low energy oscillation
%parameters. The second case is the one where we consider a quasidegenerate
%mass spectrum that can come from addition of three right handed heavy
%neutrinos degenerate in masses. Here we construct the Yukawa
%coupling using the Casasa-Ibarra parameterization \cite{Casas:2001sr}.

%First we make an analytical study of RG effect of neutrino parameters
%assuming the mass matrix at high scale is Tribimaximal (TBM). By making
%small perturbations in high energy parameters; angles and phases we get the expressions
%for parameters at low energy. Then to supplement our study quantitatively
%we perform the numerical analysis by solving the group of coupled beta functions.
%Though the beta functions are valid whether we adopt a buttom-up or top-down
%apparoach, we choose the former one i.e running from low to high scale.
%Adopting this approach, we make use of all the low energy experimental data
%available so far.

%
\begin{table}[t]\centering
  \begin{tabular}{|c|c|c|c|c|c|c|}
    \hline
 \begin{tabular}{c}
       Oscillation \\ 
       parameters
    \end{tabular} & 
    \begin{tabular}{c}
       $\Delta m^2_{21}$ \\ 
      $[10^{-5}~\text{eV}^2]$
    \end{tabular}&
    \begin{tabular}{c}
       $\Delta m^2_{31}$ \\ 
      $[10^{-3}~\text{eV}^2]$
    \end{tabular}
   & $\sin^2\theta_{12}$ &
    $\sin^2\theta_{23}$ &   $\sin^2\theta_{13}$ & $\delta$  \\
    \hline    
 Best fit values & 7.62 & 
 \begin{tabular}{c}
      2.55\\
      2.43
    \end{tabular} & 0.320 &
     \begin{tabular}{c}
      0.613\\
      0.600
    \end{tabular} & 
    \begin{tabular}{c}
      0.0246\\
      0.0250
    \end{tabular} &
    \begin{tabular}{c}
     $0.80\pi$\\
     $-0.03\pi$
   \end{tabular}  \\
    \hline    
     $ 3 \sigma$ range &  7.12--8.20 & 
    \begin{tabular}{c}
      $2.31-2.74$\\
      $2.21-2.64$
    \end{tabular} &
    0.27--0.37 & 
    \begin{tabular}{c}
      0.36--0.68\\ 
      0.37--0.67 
    \end{tabular} &  0.017--0.033 &
     $0-2\pi$ \\
       \hline
\end{tabular}
\caption{  Present best fit values and $3\sigma$ ranges of neutrino
oscillation parameters. The upper (lower) sub-row corresponds to normal
(inverted) hierarchy \cite{Tortola:2012te}. 
Values of $\Delta m^2_{21}$ and $\sin^2\theta_{12}$ 
 are hierarchy independent.}
\label{tab:oscptm}
\end{table}
The plan of the paper is as follows. 
In the next section we briefly outline the origin of neutrino masses
in type-I seesaw mechanism and present the mixing matrix and the mass spectra. 
In section \ref{sec:RG-thrshld}, we discuss about the threshold effect 
in RG evolution of neutrino mass matrix and the matching condition used
in the numerical work. In section \ref{sec:Analytic}, we perform the
analytical study and calculate the order of magnitude of RG effect on
various neutrino parameters. In section \ref{sec:Num-anlss}, we report
our results from the numerical study. Section \ref{sec:concl} summarizes
our conclusions.

\section{Neutrino Masses and Mixing} 
The Lagrangian containing the Yukawa couplings of the leptonic fields, $l_L$ 
including the heavy right handed neutrino and Higgs field, $\phi$ is
\begin{equation} 
-{\cal{L}} = \tilde{\phi}^\dagger \overline{N_R} Y_{\nu} l_L +
\phi^\dagger \overline{E_R} Y_l \,l_L + \frac{1}{2}
{\overline{N_R}} {M_R} N_R^c + \text{h.c.}
\label{eq:yyukawa} 
\end{equation} 
where $\tilde{\phi} = i \sigma^2 \phi^{*}$ , 
$E_R$ is the right handed charged 
lepton singlet and $N_R$ denotes the right handed singlet neutrino state. 
After integrating out the heavy fields, $N_R$ the above Lagrangian
gives rise to an effective dimension 5 operator 
\begin{equation}
 {\mathcal{L}_{eff}} = \frac{1}{4}\,\left(\overline{l^{{\cal{C}}}_{L}}\,\epsilon
 \, \phi\right)\kappa\left(\phi^T \epsilon^T l_{L}\right)+ \text{h.c.}
\end{equation}
where 
\begin{equation} 
\kappa = 2 Y_\nu^T {M_R}^{-1} Y_\nu
\label{seesaw1}
\end{equation} 
The mass matrix, $m_\nu$ is related to $\kappa$ as $m_\nu = (1/4)\kappa \, v^2 $
and can be diagonalized as  
\begin{eqnarray}
U^T m_{\nu} U = m_{\mathrm diag},
\end{eqnarray}
where $U$ can be identified as $U_{PMNS}$ in the basis where 
charged lepton matrices are diagonal. 
\begin{equation}
\label{eq:Upara}
 U =  \left(
 \begin{array}{ccc}
 c_{12} \, c_{13} & s_{12}\, c_{13} & s_{13}\, e^{-i \delta}\\
 -c_{23}\, s_{12}-s_{23}\, s_{13}\, c_{12}\, e^{i \delta} &
 c_{23}\, c_{12}-s_{23}\, s_{13}\, s_{12}\,
e^{i \delta} & s_{23}\, c_{13}\\
 s_{23}\, s_{12}-\, c_{23}\, s_{13}\, c_{12}\, e^{i \delta} &
 -s_{23}\, c_{12}-c_{23}\, s_{13}\, s_{12}\,
e^{i \delta} & c_{23}\, c_{13}
 \end{array}
 \right) P \, .
\end{equation}
Here $c_{ij} = \cos \theta_{ij}$, $s_{ij} = \sin \theta_{ij}$,
$\delta$ is the Dirac-type CP-violating phase
and the Majorana phases
$\alpha_1$ and $\alpha_2$ are contained in the matrix, 
$P = {\rm diag}( e^{i \alpha_1}, e^{i \alpha_2}, 1)$. While all
phases are currently unconstrained, the other mixing parameters are
determined with increasing precision \cite{Schwetz:2008er}.
The current best-fit values and 3$\sigma$ ranges of oscillation parameters
are presented in Table \ref{tab:oscptm}. Note that  
oscillation experiments can measure only the mass squared differences.
Depending upon the relative ordering of the mass states there can be 
three possible mass orderings. These are 
\begin{itemize}
\item[(i)] Normal Hierarchy (NH): in this case
$m_1 \approx m_2 < m_3 $ 
\item[(ii)] Inverted Hierarchy (IH) : this corresponds to
$m_3 < m_2 \approx m_1$. 
%Here, $\Delta m_{ij}^2 = m_j^2 - m_i^2$.
\item[(ii)] Quasidegenerate (QD) : this corresponds to
$m_3 \approx m_2 \approx m_1 >> \Delta m^2_{31}$
\end{itemize}
Currently all three possibilities are allowed although the 
QD regions is being constrained from cosmological observations \cite{Ade:2013ktc}. 
%Note that the effective dimension 5 operator is in general defined 
%at the high scale determined by the mass of the heavy neutrino. 
%The mixing angles and masses on the other hand are experimentally 
%measured values at a lower energy and in general one has to 
%include the effects of RG running of the coefficient 
%$\kappa$ leading to RG corrected
%predictions for the oscillation parameters. 
In the next section we discuss the beta functions in presence 
of heavy right-handed neutrinos coupled to the theory including 
the threshold effects.

\section{Renormalization group equation including Threshold Effect}
\label{sec:RG-thrshld}

Consider adding $q$ number of right handed heavy neutrinos of masses
$M_1...M_{q-1}, M_q$ to the SM, where $M_q$ is mass of the heaviest neutrino.
Above the scale $M_q$ the mass of the light neutrino is generated 
by Type-I seesaw mechanism and is given by 
\cite{Grzadkowski:1987tf, Einhorn:1992um, Antusch:2002rr,Antusch:2003vh, Luo:2002ey}, 
\begin{equation}
 m^{q+1}_{\nu}(\mu)= - \frac{v^2}{2} {Y^{q+1}_{\nu}}^T(\mu) M_Q^{-1}(\mu) 
 Y^{q+1}_{\nu}(\mu)
\end{equation}
where $\mu$ is the renormalization scale. $y^{(q+1)}$ is a 
$(q \times 3)$ dimensional matrix. Here $M_Q$ is a $(q \times q)$ 
complex symmetric matrix whose eigenvalues are $M_1...M_{q-1}, M_q$.
In this energy regime the neutrino masses and mixing parameters
are governed by beta function and is given by,
\begin{equation}
16 \pi^2 \frac{d m^{q+1}_{\nu}}{dt} = - \frac{1}{2} ( 3 Y^{\dagger}_e Y_e  
- {Y^{q+1}_{\nu}}^{\dagger}\,Y^{q+1}_{\nu})^T m_{\nu} - \frac{1}{2} m_{\nu} (3 Y^{\dagger}_e Y_e -
{Y^{q+1}_{\nu}}^{\dagger}\,Y^{q+1}_{\nu}) + \alpha_m m_{\nu}
\label{eq:RG-eqn}
\end{equation}
where 
\begin{equation}
\alpha_m = 2 {\rm Tr} ( Y^{\dagger}_e Y_e + Y^{\dagger}_{\nu} Y_{\nu})
 + 6 {\rm Tr} (Y^{\dagger}_u Y_u + Y^{\dagger}_d Y_d)
- \frac{9}{2} g^2_2 - \frac{9}{10} g_1^2 \nonumber
\end{equation}
At the energy scale $M_q$ the heaviest neutrino gets integrated
out and produces dimension five effective operator
given as
%\begin{equation}
% {\mathcal{L}_{eff}} = \frac{1}{4}\kappa l \, l\, \phi\, \phi
%\end{equation}
%where the coefficient $\kappa$ is given by
\begin{equation}
 \kappa^q_{ij}= 2 ({Y^{q+1}}^T_{\nu})_{iq} (M_q^{-1}) (Y^{q+1}_{\nu})_{qj} \,\,
 \,( {\rm no \, sum \, over\,} q)
\label{eq:matching-cond}
\end{equation}
The above equation can be interpreted as the continuity of the
neutrino mass matrix at the scale $\mu = M_q$. Hence, the
mixing parameters at the particular energy scale can be extracted
using standard procedure \cite{Antusch:2003kp}. Now consider the regime where
$M_{q-1} < \mu < M_q$. Here the neutrino mass gets contributions from
two sources, one from the standard seesaw mechanism due to presence
of heavy neutrinos and the
second from the effective operator $\kappa$ and is  given by
\begin{equation}
 m^q_{\nu} = -\frac{v^2}{4} (\kappa^q + 2 {Y^q_{\nu}}^T M_{Q-1}^{-1} Y^q_{\nu})
\label{eq:nuMAss-kappa-seesaw}
\end{equation}
where $y^{(q)}$ is a $((q-1) \times 3 )$ dimensional matrix. $Y^q_{\nu}$
is obtained by setting the elements of $q^{\rm th}$ row of $Y_{\nu}$ as
zero in the basis where $M_Q$ is diagonal. $M_{Q-1}$ is obtained by
removing the last row and last column of $M_Q$.
At this stage the RG effects again come into picture. The beta function
for the second term is same as given in Eq.(\ref{eq:RG-eqn}) with
$Y_{\nu}$ replaced by $Y^q_{\nu}$. However
the equation for $\kappa$ is given by \cite{Antusch:2005gp}
\begin{equation}
16 \pi^2 \frac{d\kappa^q}{dt} 
= -\frac{1}{2}(h^2_i \kappa^q + \kappa^q h^2_j) + 2 h^2_k \kappa^q
+ 6 h^2_l \kappa^q + h^2_m \kappa^q
\label{eq:RG-eqn1}
\end{equation}
where 
\begin{eqnarray}
 h^2_i &=& 3 ( Y^{\dagger}_e Y_e )^T - ({Y^q_{\nu}}^{\dagger}\,Y^q_{\nu})^T \nonumber \\
 h^2_j &=& 3 ( Y^{\dagger}_e Y_e ) - ({Y^q_{\nu}}^{\dagger}\,Y^q_{\nu}) \\ \nonumber
 h^2_k &=&  {\rm Tr} ( Y^{\dagger}_e Y_e ) + {\rm Tr} ({Y^q_{\nu}}^{\dagger}\,Y^q_{\nu}) \nonumber \\
 h^2_l &=&  {\rm Tr} ( Y^{\dagger}_u Y_u ) + {\rm Tr} (Y^{\dagger}_d Y_d)  \nonumber \\
 h^2_m  &=& \lambda_5 - 3g_2^2
\label{eq:h-squares}
\end{eqnarray}
%\begin{equation}
%16 \pi^2 \frac{d\kappa}{dt} = - \frac{1}{2} (3 Y^{\dagger}_e Y_e-
%Y^{\dagger}_{\nu} Y_{\nu})^T - \frac{1}{2}  (3 Y^{\dagger}_e Y_e-
%Y^{\dagger}_{\nu} Y_{\nu}) (Y^{\dagger}_u Y_u) \kappa + \alpha_{\kappa} \kappa
%\label{eq:RG-eqn2}
%\end{equation}
% where 
%\begin{equation}
% \alpha_{\kappa} = 2 {\rm Tr} ( Y^{\dagger}_e Y_e + Y^{\dagger}_{\nu} Y_{\nu})
% + 6 {\rm Tr} (Y^{\dagger}_u Y_u +Y^{\dagger}_d Y_d) 
%- 3 g^2_2 + \lambda 
%\end{equation}
The beta functions for the Yukawa couplings for leptons and neutrinos are given
by
\begin{eqnarray}
 16 \pi^2 \frac{d Y_{\nu}}{d t} &=& Y_{\nu}\left\{ \frac{3}{2}
 (Y^{\dagger}_{\nu} Y_{\nu}) - \frac{3}{2}  ( Y^{\dagger}_e Y_e )
 +  {\rm Tr} (Y^{\dagger}_{\nu} Y_{\nu})+ {\rm Tr} (Y^{\dagger}_e Y_e)\right\} \nonumber \\   
& &+ Y_{\nu}\left\{ 3 {\rm Tr} (Y^{\dagger}_u Y_u) + 3 (Y^{\dagger}_d Y_d) 
 - \frac{9}{20} g_1^2 -  \frac{9}{4} g_2^2 \right\}
\label{eqn:beta-ynu}
\end{eqnarray}
\begin{eqnarray}
 16 \pi^2 \frac{d Y_e}{d t} &=& Y_e \left\{ \frac{3}{2}  ( Y^{\dagger}_e Y_e )
-\frac{3}{2}  (Y^{\dagger}_{\nu} Y_{\nu})+ 
{\rm Tr} (Y^{\dagger}_{\nu} Y_{\nu}+Y^{\dagger}_e Y_e)\right\} \nonumber    \\
&+&Y_e \left\{ 3 {\rm Tr} (Y^{\dagger}_u Y_u + Y^{\dagger}_d Y_d) -
 \frac{9}{4} g_1^2 -  \frac{9}{4} g_2^2\right\}
\end{eqnarray}
The Majorana mass scale, $M_R$ is also governed by beta function,
\begin{equation}
 16 \pi^2 \frac{d M_R}{d t} = (Y_{\nu} Y^{\dagger}_{\nu}) M_R +
M_R (Y^{\dagger}_{\nu} Y_{\nu})^T
\end{equation}
 
Proceeding as above, integrating out the heavy Majorana neutrino
fields sequentially while taking care of matching conditions
at particular decoupling point one reaches the effective 
theory \cite{Bergstrom:2010id, Bergstrom:2010qb,Antusch:2005gp}.

%In this particular work we have added two extra fields; one singlet
%Majorana fermion and a gauge singlet having opposite lepton numbers.
%We consider a low scale seesaw mechanism where the decoupling of
%extra degrees of freedom takes place at energy scale $M_R \simeq 10^3$GeV.
%Above the scale $M_R$ the RG evolution of neutrino mass matrix is
%as given on Eq.(\ref{eq:RG-eqn}). Below the scale $M_R$ it is the effective
%theory and Eq.(\ref{eq:RG-eqn1}) takes care of the evolution of
%neutrino mass matrix. The matching condition at $M_R$ scale is given in 
%Eq.(\ref{eq:matching-cond}).
%
\section{Analytical Results}
\label{sec:Analytic}
The running of neutrino mass matrix can be obtained  using the
prescription in \cite{Ellis:1999my,Chankowski:2001mx},
\begin{equation}
 M^{\lambda}_{\nu} = I_S\cdot I^T \cdot M^{\Lambda}_{\nu} \cdot I
 \label{eq:factorised}
\end{equation}
where $M^{\Lambda}_{\nu}$ and $M^{\lambda}_{\nu}$ are neutrino mass matrices 
at high and low scale respectively. $I_S $ is a factor that arises due to
RG evolution of gauge coupling constants, Higgs self coupling and
fermion Yukawa couplings and is given by
\begin{equation}
 I_S = {\rm Exp} \left\{\frac{1}{16 \pi^2} \int^t_{t_0}  (h^2_k + h^2_l
 +  h^2_m ) \right\}
 \label{eq:IS-eqn}
\end{equation}
The matrix $I$ appearing in Eq.(\ref{eq:factorised}) is given by
\begin{equation}
 I = 
\begin{pmatrix}
 \sqrt{I_e} & 0 & 0\\
0 & \sqrt{I_{\mu}} & 0 \\
0 & 0 & \sqrt{I_{\tau}}
\end{pmatrix}
\end{equation}
where for the present case
\begin{eqnarray}
\sqrt{I_j} &=& {\rm Exp} \left\{- \frac{1}{16 \pi^2} \int 
(3 ( Y^{\dagger}_j Y_j ) - (Y^{\dagger}_{\nu_j} Y_{\nu_j})) dt \right\} =
 e^{-\Delta_j},~~ j =e, \mu, \tau
% \\ \nonumber 
%\sqrt{I_{\mu}} &=& {\rm Exp} \left\{- \frac{1}{16 \pi^2} \int 
%(3 ( Y^{\dagger}_{\mu} Y_{\mu} ) - (Y^{\dagger}_{\nu_{\mu}} Y_{\nu_{\mu}})) dt  \right\}
%=   e^{-\Delta_{\mu}} \\ \nonumber 
%\sqrt{I_{\tau}} &=&  {\rm Exp} \left\{- \frac{1}{16 \pi^2} \int 
%(3 ( Y^{\dagger}_{\tau} Y_{\tau} ) - (Y^{\dagger}_{\nu_{\tau}} Y_{\nu_{\tau}})) dt  \right\} 
%= e^{-\Delta_{\tau}}
\end{eqnarray}
%In another way $I$ can be written as 
%\begin{equation}
% I = {\rm Diag} (e^{-\Delta_e}, e^{-\Delta_{\mu}}, e^{-\Delta_{\tau}})
%\end{equation}
where
%\begin{equation}
% \Delta_i= \frac{1}{16 \pi^2} \int 
%(3 ( Y^{\dagger}_i Y_i ) - (Y^{\dagger}_{\nu_i} Y_{\nu_i})) dt, i = e, \mu, \tau
%\end{equation}
where $t =  {\rm ln} (Q/Q_0)$, $Q$ being the running scale and $Q_0$ is the
fixed scale. We can calculate the order of magnitude of $\Delta_j$; 
for example, $\Delta_{\tau}$. For $Y_{\tau} \sim 10^{-2}, Y_{\nu_\tau} \sim 0.2$,
$\Lambda= 10^{12}$ GeV and $\lambda = 10^2$ GeV, $\Lambda$ ($\lambda$) 
being the high (low) scales respectively, we get 
\begin{equation}
 |\Delta_{\tau}| \sim 5.2 \times 10^{-3}
\label{Delta-tau-approx}
\end{equation}
For the sake of comparison the value of $\Delta_{\tau}$ can
be calculated where threshold effect due to neutrinos are absent
i.e. there are no $Y_{\nu}$ terms in the beta functions. In this case
\begin{equation}
 | \Delta_{\tau}|_{{\rm without }\,Y_{\nu}}\sim 3.9 \times 10^{-5}
\label{Delta-tau-approx-wthoutYnu}
\end{equation}
For this case the order of magnitude for $\Delta_e$ and $\Delta_{\mu}$
can be calculated to be $\sim 10^{-11}$ and $\sim 10^{-7}$ respectively
which are negligible in comparison to $\Delta_{\tau}$. But in the case where we
include threshold effects  $\Delta_e$ and $\Delta_{\mu}$ can be comparable
to $\Delta_{\tau}$.
Thus including threshold corrections the running is expected to
 be more. However 
$\Delta_i$ is still small to allow for a linear approximation and 
the mass matrix  at a lower scale $\lambda$ can be written as,  
\begin{equation}
 M^{\lambda}_{\nu} = I_S  
\begin{pmatrix}
 1-\Delta_e & 0 & 0\\
0 & 1-\Delta_{\mu} & 0\\
0 & 0 & 1-\Delta_{\tau}
\end{pmatrix} \,
M^{\Lambda}_{\nu} \,
\begin{pmatrix}
 1-\Delta_e & 0 & 0\\
0 & 1-\Delta_{\mu} & 0\\
0 & 0 & 1-\Delta_{\tau}
\end{pmatrix}
\end{equation}
 Since the neutrino mass matrices at low scale as well as
at high scale are complex symmetric, they can be diagonalized as
\begin{eqnarray}
(U^{\lambda})^T M^{\lambda}_{\nu} U^{\lambda} &=& 
{\rm Diag} (|m^{\lambda}_1|, |m^{\lambda}_2|, |m^{\lambda}_2|) \nonumber \\
(U^{\Lambda})^T M^{\Lambda}_{\nu} U^{\Lambda} &=& 
{\rm Diag} (|m^{\Lambda}_1|, |m^{\Lambda}_2|, |m^{\Lambda}_3|)
\label{eq:Diagonalisation}
\end{eqnarray}
%The information about mixing angles and phases are only contained 
%in the mixing matrix $U$. Whereas the informmovie on antimatteration about masses 
%only appear in the effctive operator $\kappa$. 
$U^{\Lambda}$ and $U^{\lambda}$ are diagonalizing matrices containing the unphysical leptonic phases, $\Phi_i$.
The mixing angles and phases at low scale are related to that at high scale 
up to first order in $\Delta_e \,, \Delta_\mu \, \text{and}\, \Delta_\tau$  as
\begin{equation}
\theta^{\lambda}_{ij} = \theta^{\Lambda}_{ij} + K_{e_{ij}}\Delta_e +
K_{\mu_{ij}}\Delta_{\mu} + K_{\tau_{ij}}\Delta_{\tau} ; \, i, j = 1,2,3
\label{eq:angles-low-high}
\end{equation}
%The Dirac and Majorana phases at low scale are related to that at high scale as
\begin{eqnarray}
\delta^{\lambda} &=& \delta^{\Lambda} + d_e \Delta_e+
d_{\mu} \Delta_{\mu} +  d_{\tau}\Delta_{\tau}  \nonumber \\
\alpha^{\lambda}_i &=& \alpha^{\Lambda}_i+  a_{e_i}\Delta_e +
 a_{\mu_i} \Delta_{\mu} +  a_{\tau_i}\Delta_{\tau}; \, i = 1,2
\label{eq:majorana-Dirac-low-high}
\end{eqnarray}
%
% The leptonic phases at high scale to that at low scale can be
% related as,
%
\begin{equation}
  \Phi_j^{\lambda} = \Phi_j^{\Lambda}+  p_{e_j}\Delta_e +
 p_{\mu_j} \Delta_{\mu} + p_{\tau_j}\Delta_{\tau} ; \, j = 1,2 ,3
 \label{eq:leptonic-low-high}
\end{equation}
It is possible to obtain analytic expressions for the $K_{ij}$'s, 
$a_i$'s, $d$'s and $p_i$'s in the limit of small $\theta_{13}$ keeping
$\sin \theta_{13}$ up to second order. The expressions involved are quite 
long  and are given in \cite{Bergstrom:2010qb} for the 
masses, mixing angles and the Dirac CP phase.  These match with what 
we obtain using the method outlined in this paper.   
Thus we do not give these expressions  
again in this paper. However in the appendix we give  the coefficients 
involved in the running of the Majorana as well as the leptonic phases, which 
were not given in \cite{Bergstrom:2010qb}. 
To gain some analytic understanding of the running, below we give the 
expressions for the coefficients 
in the limit $\theta_{13}=0$  as, 
\begin{eqnarray}
 K_{e_{12}} &=& -\frac{1}{2} {\rm sin}\, 2 \theta^{\Lambda}_{12}\,
\frac{|m^{\Lambda}_1 + m^{\Lambda}_2|^2}{|m^{\Lambda}_2|^2 - |m^{\Lambda}_1|^2} \nonumber  \\
K_{\mu_{12}} &=& -  {\rm cos}^2\theta^{\Lambda}_{23} 
K_{e_{12}}\nonumber \\ 
 K_{\tau_{12}} &=& -  {\rm sin}^ 2\theta^{\Lambda}_{23} K_{e_{12}}\nonumber  \\ 
 K_{e_{13}} &=& 0 \\ \nonumber
K_{\mu_{13}} &=& -\frac{1}{4} {\rm sin}\, 2 \theta^{\Lambda}_{12}\,{\rm sin}\, 2 \theta^{\Lambda}_{23}\,
\left(\frac{|m^{\Lambda}_2 + m^{\Lambda}_3|^2}{|m^{\Lambda}_3|^2 - |m^{\Lambda}_2|^2} -
\frac{|m^{\Lambda}_3 + m^{\Lambda}_1|^2}{|m^{\Lambda}_3|^2 - |m^{\Lambda}_1|^2}\right) \nonumber \\
K_{\tau_{13}} &=& -K_{\mu_{13}} \\ \nonumber
 K_{e_{23}} &=& 0 \nonumber  \\
K_{\mu_{23}} &=& -\frac{1}{2} {\rm sin}\, 2 \theta^{\Lambda}_{23}\,
\left(\frac{|m^{\Lambda}_2 + m^{\Lambda}_3|^2}{|m^{\Lambda}_3|^2 - |m^{\Lambda}_2|^2}
{\rm cos}^2 \theta^{\Lambda}_{12} +
\frac{|m^{\Lambda}_3 + m^{\Lambda}_1|^2}{|m^{\Lambda}_3|^2 - |m^{\Lambda}_1|^2}
{\rm sin}^2 \theta^{\Lambda}_{12}\right) \nonumber  \\
K_{\tau_{23}} &=& - K_{\mu_{23}}
\label{eq:Kij-eqns}
\end{eqnarray}
where
\begin{equation}
 m_1^\Lambda = |m_1^\Lambda| e^{i\alpha_1^\Lambda},  
 m_2^\Lambda = |m_2^\Lambda| e^{i\alpha_2^\Lambda}, 
 m_3^\Lambda = |m_3^\Lambda| 
\label{eq:Masses}
\end{equation}
In the limit $\theta^{\Lambda}_{13}=0$ 
the $a_i$'s can be written as
\begin{eqnarray}
 a_{e1} &=& \frac{4 |m^{\Lambda}_1| |m^{\Lambda}_2|}{|m^{\Lambda}_2|^2-
 |m^{\Lambda}_1|^2} \sin(\alpha^{\Lambda}_1 -\alpha^{\Lambda}_2) 
 \cos^2\theta^{\Lambda}_{12} \nonumber  \\
 a_{e2} &=& a_{e1}\tan^2\theta^{\Lambda}_{12} \nonumber  \\
 a_{\mu 1} &=& - a_{e1}  \cos^2 \theta^{\Lambda}_{23} \nonumber  \\
 a_{\mu 2} &=& - a_{e2}  \cos^2 \theta^{\Lambda}_{23} \nonumber  \\
 a_{\tau 1} &=& - a_{e1}  \sin^2 \theta^{\Lambda}_{23} \nonumber  \\
 a_{\tau 2} &=& - a_{e2}  \sin^2 \theta^{\Lambda}_{23}
 \label{eq:a-i-eqns}
\end{eqnarray}
The analytic expressions encoding the evolution of Dirac CP phase is given
in Appendix(\ref{sec:app}) in terms of $d_e, d_{\mu}$ and $d_{\tau}$.
We can write a generalize expression for the later as
\begin{equation}
 d_i = \frac{A_{i}}{\sin \theta^{\Lambda}_{13}} + B_{i} +C_{i} \sin \theta^{\Lambda}_{13} 
 + O (\sin^2 \theta^{\Lambda}_{13})
 \label{eq:V-i}
\end{equation}
where $i = e, \mu, \tau$.
It is easy to see from the expressions given in the appendix that 
$A_{e} = 0$ whereas 
$A_{\mu}$ and $A_{\tau}$ are exactly equal and opposite to each other.
At the first place it appears that first term in
Eq.(\ref{eq:V-i}) will diverge for vanishing $\theta^{\Lambda}_{13}$ resulting
in the discontinuity in the running of Dirac CP phase $\delta$. Such a 
behavior is anomalous since all the neutrino parameters evolve continuously
with renormalization scale. 
%However, from the expression
%It is easy to see that $v_{e1} = 0$ whereas 
%$v_{\mu1}$ and $v_{\tau1}$ are exacly equal and opposite to each other.
In order to ensure the continuity of running of $\delta$, one  imposes the 
condition 
$A_{\mu} = A_{\tau} = 0 $ which fixes the CP phase $\delta$ as  
\begin{equation}
 \cot \delta = \frac{m^{\Lambda}_1 \cos \alpha^{\Lambda}_1 - 
 m^{\Lambda}_2 (1+r) \cos \alpha^{\Lambda}_2 - r m^{\Lambda}_3}
 {m^{\Lambda}_1 \sin \alpha^{\Lambda}_1 - m^{\Lambda}_2 (1+r) \sin \alpha^{\Lambda}_2 }
\end{equation}
Note that even after including the threshold effects this condition 
remains same as in \cite{Antusch:2003kp} since $A_{\mu} = -A_{\tau}$. 
A more general discussion can be found in \cite{Dighe:2008wn}.

The masses at low scale to that at high scale are related as
\begin{eqnarray}
 |m_1^{\lambda}| &=& I_S\, |m_1^{\Lambda}| (1-2 \Delta_e {\rm cos}^2 \theta^{\Lambda}_{12}
-2 \Delta_{\mu} {\rm sin}^2\theta^{\Lambda}_{12} {\rm cos}^2 \theta^{\Lambda}_{23}
-2\Delta_{\tau} {\rm sin}^2\theta^{\Lambda}_{12} {\rm sin}^2 \theta^{\Lambda}_{23} ) \nonumber \\
|m_2^{\lambda}| &=& I_S \, |m_2^{\Lambda}| (1-2\Delta_e {\rm cos}^2 \theta^{\Lambda}_{12}
-2\Delta_{\mu} {\rm cos}^2 \theta^{\Lambda}_{12} {\rm cos}^2 \theta^{\Lambda}_{23}
-2 \Delta_{\tau} {\rm cos}^2 \theta^{\Lambda}_{12}{\rm sin}^2 \theta^{\Lambda}_{23} ) \nonumber \\
|m_3^{\lambda}| &=& I_S \,|m_3^{\Lambda}| (1-2\Delta_{\mu} {\rm sin}^2 \theta^{\Lambda}_{23} 
- 2\Delta_{\tau} {\rm cos}^2 \theta^{\Lambda}_{23})
\label{eq:mass-exp}
\end{eqnarray}
%
% To get an order of maginutude estimate of running of angles
% consider the case of tri-bimaximal (TBM) mixing at the high scale such that 
% $\theta^{\Lambda}_{12}={\rm sin}^{-1}\sqrt{1/3} $,
% $\theta^{\Lambda}_{23}= \pi/4$, and $\theta^{\Lambda}_{13}=0$.
% In the limit $k_{ij} \Delta_i \ll 1$, the effect of running
% of the angles can be seen as
% 
% \begin{eqnarray}
% \theta^{\lambda}_{12} &\simeq& {\rm sin}^{-1} \sqrt{1/3}  + 
% K_{e_{12}}\Delta_e + K_{\mu_{12}}\Delta_{\mu} + K_{\tau_{12}} \Delta_{\tau} \nonumber  \\
% \theta^{\lambda}_{23} &\simeq&\frac{\pi}{4} + K_{e_{23}}\Delta_e +
% K_{\mu_{23}}\Delta_{\mu} + K_{\tau_{23}} \Delta_{\tau} \nonumber  \\
% \theta^{\lambda}_{13} &\simeq&  K_{e_{13}}\Delta_e +
% K_{\mu_{13}}\Delta_{\mu} + K_{\tau_{13}} \Delta_{\tau}
% \label{eq:angles-effctv}
% \end{eqnarray}
%
For non zero value of $\theta_{13}^{\Lambda}$, the $K_{ij}$'s have an 
error of $\mathcal{O}\left(\theta_{13}^{\Lambda}\right)$ and the mixing 
angles and phases will have error $\mathcal{O}\left(\theta_{13}^{\Lambda}\,\Delta_e\right)$. 
Also in order that $\mathcal{O}\left(\Delta_e^2\right)$ terms do not dominate 
over $\mathcal{O}\left(\Delta_e\right)$ terms, one needs 
\begin{eqnarray}
m^2_0 \left( 1 + \cos (\alpha^{\Lambda}_1 -\alpha^{\Lambda}_2)\right)\Delta_e < |{m^{\Lambda}_2}^2| -|{m^{\Lambda}_1}^2|
\label{eq:k-validity}
\end{eqnarray}
Similarly the validity of $a_i$'s and $d$'s requires
\begin{eqnarray}
m^2_0 \sin (\alpha^{\Lambda}_1 -\alpha^{\Lambda}_2) \Delta_e < |{m^{\Lambda}_2}^2| -|{m^{\Lambda}_1}^2|
\label{eq:a-validity}
\end{eqnarray}

From Eq.(\ref{eq:angles-low-high}) it can be stated as
\begin{eqnarray}
 \sin^2 \theta^{\lambda}_{12}- \sin^2 \theta^{\Lambda}_{12}&\simeq&
  K_{e_{12}} \left( \Delta_e - 
 \cos^2  \theta^{\Lambda}_{23} \Delta_{\mu} -
 \sin^2  \theta^{\Lambda}_{23} \Delta_{\tau} \right) \nonumber  \\
  \sin^2 \theta^{\lambda}_{23}-\sin^2 \theta^{\Lambda}_{23} &\simeq& 
  K_{\mu_{23}} \left( \Delta_{\mu} - \Delta_{\tau} \right)\nonumber  \\
  \sin^2 \theta^{\lambda}_{13}-\sin^2 \theta^{\Lambda}_{13} &\simeq& 
  K_{\mu_{13}} \left( \Delta_{\mu} - \Delta_{\tau} \right)
  \label{eq:sin-sqr-thetas}
\end{eqnarray}

%the Eqs. (\ref{eq:Kij-eqn})

The expressions for the $K_{ij}$'s given  
in Eqs. (\ref{eq:Kij-eqns})
can be further simplified depending on the mass 
spectrum. Below we give the expressions for NH, IH and QD cases. 
The dependence on the Majorana phases become apparent from these 
expressions. 
%We make use of one parameter, $r$ as
%\begin{equation}
% r = \frac{\Delta m^2_{\rm sol}}{\Delta m^2_{\rm atm}}
% \label{eq:r-eqn}
%\end{equation}

\begin{itemize}
\item  {Normal hierarchy}

For NH one can use the approximation,
$m_1 \approx 0, m_2  \simeq \sqrt{\Delta m^2_{21}}$
and $m_3  \simeq \sqrt{\Delta m^2_{32}(1+r)}$, 
where $r = \Delta m^2_{21}/\Delta m^2_{32}$ 
Using the above one can express the $K_{ij}$'s for NH as,  
\begin{eqnarray}
K_{e_{12}} & \approx & -\frac{1}{2} \sin2\theta_{12}^\Lambda \nonumber \\
%K_{\mu_{12}} & \approx & \frac{1}{2} \cos^2\theta_{23}^\Lambda \sin2\theta_{12}^\Lambda \\
%K_{\tau_{12}} & \approx  & \frac{1}{2} \sin^2\theta_{23}^\Lambda \sin 2\theta_{12}^\Lambda
K_{\mu_{13}} & \approx & -\frac{1}{2} \sin2\theta_{12}^\Lambda \sin2\theta_{23}^\Lambda 
[ r + \sqrt{r} \cos(\alpha^\Lambda_2) ] \nonumber \\
K_{\mu_{23}} & \approx & -\frac{1}{2} \sin2\theta_{23}^\Lambda [ 
1+2 \cos^2\theta^{\Lambda}_{12}(r + \sqrt{r} \cos(\alpha^\Lambda_2)) ]
\label{eq:kij-NH}
\end{eqnarray} 
\item {Inverted hierarchy} 

For IH one can make the simplification, $r = \Delta m^2_{21}/ \Delta m^2_{13}$,
$m_3 \approx 0$,$m_1 \approx \sqrt{\Delta m^2_{13}+ m_3^2}$,
$m_2 \approx \sqrt{\Delta m^2_{13}(1+r)+m_3^2}$.
Then one can write, 
\begin{eqnarray}
K_{e_{12}} &\approx & - \sin2\theta_{12}^\Lambda
\left(\frac{1+ \cos(\alpha^\Lambda_1 -\alpha^\Lambda_2)}{r}\right) \nonumber \\
K_{\mu_{13}} & \approx& 
 \frac{1}{4} \sin2\theta_{12}^\Lambda \sin2\theta_{23}^\Lambda   
\frac{m_3}{\sqrt{\Delta m^2_{13}}}[\cos\alpha^\Lambda_2 - \cos \alpha^\Lambda_1] \nonumber \\
K_{\mu_{23}} & \approx & \frac{1}{2} \sin2\theta_{23}^\Lambda [ 
1+ \frac{2m_3}{\sqrt{\Delta m^2_{13}}} (\cos\alpha^\Lambda_2
\cos^2 \theta^{\Lambda}_{12} + \cos \alpha^\Lambda_1 \sin^2 \theta^{\Lambda}_{12}) ] 
\label{eq:kij-IH}
%\cos(\alpha_2  {c_{12}^2}^\Lambda - {s_{12}^2}^\Lambda ]
\end{eqnarray}
\item {Quasidegenerate} 

In this case the expressions for the $K_{ij}$'s simplifies to
\begin{eqnarray}
K_{e_{12}} & \approx & -\sin2\theta_{12}^\Lambda \frac{{m_0}^2}{\Delta m^2_{21}} 
[ 1 + \cos(\alpha^\Lambda_1 - \alpha^\Lambda_2) ] \nonumber
\\
%K_{\mu_{12}} & \approx & \frac{1}{2} \cos^2\theta_{23}^\Lambda \sin2\theta_{12}^\Lambda \\
%K_{\tau_{12}} & \approx  & \frac{1}{2} \sin^2\theta_{23}^\Lambda \sin 2\theta_{12}^\Lambda
%\\
K_{\mu_{13}} & \approx & -\frac{1}{2} \sin2\theta_{12}^\Lambda 
\sin2\theta_{23}^\Lambda \frac{m_0^2}{\Delta m^2_{32}} [ \cos\alpha^\Lambda_2 - \cos\alpha^\Lambda_1 ]
\nonumber \\
K_{\mu_{23}} & \approx & -\frac{m_0^2}{\Delta m^2_{32}} \sin2\theta^\Lambda_{23} [ 
1 + \cos\alpha^\Lambda_2 \cos^2 \theta^\Lambda_{12} + \cos \alpha^\Lambda_1 \sin^2 \theta^\Lambda_{12} ]
\label{eq:kij-QD}
\end{eqnarray} 

\end{itemize}

From the above expressions we see that in case of NH the running 
of the angle $\theta_{12}$ does not depend on the Majorana phases.
On the other hand 
the running of the mixing angle $\theta_{13}$ and $\theta_{23}$ is
maximum when $\alpha_2 = 0$. 
For the IH and QD case running of all the angles depend on the Majorana
phases. 

% From Eq.(\ref{eq:angles-low-high}) one can write, 
% \begin{eqnarray}
%  \sin^2 \theta^{\lambda}_{12}- \sin^2 \theta^{\Lambda}_{12}&\simeq&
%   K_{e_{12}} \left( \Delta_e - 
%  \cos^2  \theta^{\Lambda}_{23} \Delta_{\mu} -
%  \sin^2  \theta^{\Lambda}_{23} \Delta_{\tau} \right) \nonumber  \\
%   \sin^2 \theta^{\lambda}_{23}-\sin^2 \theta^{\Lambda}_{23} &\simeq& 
%   K_{\mu_{23}} \left( \Delta_{\mu} - \Delta_{\tau} \right)\nonumber  \\
%   \sin^2 \theta^{\lambda}_{13}-\sin^2 \theta^{\Lambda}_{13} &\simeq& 
%   K_{\mu_{13}} \left( \Delta_{\mu} - \Delta_{\tau} \right)
% \end{eqnarray}
To get an order of magnitude estimate of running of angles
consider the case of Tri-bimaximal (TBM) mixing at the high scale such that 
$\theta^{\Lambda}_{12}={\rm sin}^{-1}\sqrt{1/3} $,
$\theta^{\Lambda}_{23}= \pi/4$, and $\theta^{\Lambda}_{13}=0$.

In that case
from Eq.(\ref{eq:sin-sqr-thetas}) one gets for NH, 
\begin{equation}
 |\sin^2 \theta^{\lambda}_{12} -\frac{1}{3}|\simeq 1.5\times10^{-4}
%{\rm sin}^{-1} \sqrt{1/3} + Ke_{12}\Delta_e +
%%\simeq 0.61
 \end{equation}
\begin{equation}
 |\sin^2 \theta^{\lambda}_{23} - \frac{1}{2}| \simeq 3.1\times 10^{-3}
% \frac{\pi}{4} + Ke_{23}\Delta_e +breaking brackets with eqnarray in latex
%K{\mu}_{23}\Delta_{\mu}+ K{\tau}_{23} \Delta_{\tau}
%\simeq 0.78
\end{equation}
\begin{equation}
|\sin^2 \theta^{\lambda}_{13}| \simeq 5.0 \times 10^{-4}
%Ke_{23}\Delta_e +
%K{\mu}_{23}\Delta_{\mu} + K{\tau}_{13} \Delta_{\tau}\simeq 0.0002
\end{equation}

%{\bf I have derived the expressions keeping the phase with 
%$m_2$ and $m_3$. What is the convention followed in the numerical work ? } 

%\subsection{Inverted hierarchy}
%
For IH one gets, 
\begin{equation}
|\sin^2 \theta^{\lambda}_{12} -\frac{1}{3}|\simeq 0.3
%  \theta^{\lambda}_{23} \simeq \frac{\pi}{4} + Ke_{23}\Delta_e +
% K{\mu}_{23}\Delta_{\mu}+ K{\tau}_{23} \Delta_{\tau}
% \simeq 0.78
\end{equation}
\begin{equation}
|\sin^2 \theta^{\lambda}_{23} - \frac{1}{2}| \simeq 2.5 \times 10^{-3} 
% \theta^{\lambda}_{12} \simeq {\rm sin}^{-1} \sqrt{1/3} +Ke_{12}\Delta_e +
% K{\mu}_{12}\Delta_{\mu}+ K{\tau}_{12} \Delta_{\tau}
% \simeq 0.8
\end{equation}
\begin{equation}
|\sin^2 \theta^{\lambda}_{13}| \simeq 4.7 \times 10^{-5}
%  \theta^{\lambda}_{13} \simeq Ke_{23}\Delta_e +
% K{\mu}_{23}\Delta_{\mu} +K{\tau}_{13} \Delta_{\tau}\simeq 0
\end{equation}
%
%\subsection{Quasidegenrate}
whereas for the QD case. 

\begin{equation}
|\sin^2 \theta^{\lambda}_{12} -\frac{1}{3}|\simeq 0.6
%  \theta^{\lambda}_{12} \simeq {\rm sin}^{-1} \sqrt{1/3} + K{\tau}_{12} \Delta_{\tau}
% \simeq 0.84
\end{equation}
\begin{equation}
|\sin^2 \theta^{\lambda}_{23} - \frac{1}{2}| \simeq 0.13
%  \theta^{\lambda}_{23} \simeq \frac{\pi}{4} + K{\tau}_{23} \Delta_{\tau}
% \simeq 0.78
\end{equation}
\begin{equation}
|\sin^2 \theta^{\lambda}_{13}| \simeq 0.07
% \theta^{\lambda}_{13} \simeq K{\tau}_{13} \Delta_{\tau}\simeq 0
\end{equation}
The estimates are obtained choosing values of phases so
as to obtain maximal running and using Eq.(\ref{Delta-tau-approx}) for $\Delta_{\tau}$
and setting $\Delta_e$ and $\Delta_{\mu}$ to zero. 
The running is seen to be maximum for the mixing angle $\theta_{12}$. 
Also, for  the same value of $\Delta_{\tau}$
the running is maximum for the QD case.
However the realistic running would require the knowledge of 
the matrix $Y_\nu$. In the next section we present the numerical analysis 
using two models for which $Y_\nu$ can be easily reconstructed.  

\section{Numerical Analysis}
\label{sec:Num-anlss}
In this section we present the results obtained by
solving the set of beta functions numerically to study
the evolution of neutrino mass parameters; the mass eigenvalues,
mixing angles and the phases. We perform the running 
%the set of coupled
%equations
from low to high scale, thus making use of all the
experimental results available at low energy. We vary the
renormalization scale from $M_Z$ (mass of Z boson) scale
to the high scale $10^{12}$ GeV.
We have taken the value of the Higgs mass $m_h = 126.6$ GeV. 
The threshold correction for top quark mass contribution
is taken care of. We consider  the specific case of the Minimal Linear
Seesaw Model (MLSM) for which the Yukawa coupling
matrices are reconstructible from low energy oscillation 
parameters apart from an overall constant 
\cite{Gavela:2009cd,Khan:2012zw}. In this model one of the
mass eigenvalues is zero and thus this gives hierarchical neutrinos. 
We also consider the case of quasidegenerate neutrinos and use the 
Casas-Ibarra parametrization to determine the Yukawa texture.  
%The values of the Yukawa coupling constants
%are reconstructed from oscillation parameters.
We have used
the experimental values available for the mass eigenvalues and
mixing angles, hence exploiting all the low energy data available
so far. The phases have been varied randomly covering their full range. 

To solve the beta functions we start with the effective theory, which
is the SM in our case. 
For a specific value of the lowest mass ( $m_1$ for NH or $m_3$ for IH) 
the other two masses are  fixed in terms of the mass squared differences 
measured in oscillation experiments. 
We first compute
$\kappa$ using the low scale masses and mixing angles
and then run it up to the TeV scale. 
At TeV scale we impose the matching condition given in
Eq.(\ref{eq:matching-cond}) and extract the Yukawa matrices 
$Y_\nu$. 
We adopt the procedure of running and diagonalizing
at each step of the iteration. It should be kept in mind that while 
running the neutrino mass matrix elements they are liable to mix,
hence it is required to diagonalize them at each step. The neutrino
parameters are extracted by using the standard procedure given in
in \cite{Antusch:2003kp}. At the coupling/decoupling scale we use 
the matching condition
given in Eq.(\ref{eq:matching-cond}). Then we consider the running
of the parameters of the full theory. 
%We run the coupled equations
%for both mass ordering; Normal and inverted.

\subsection{Hierarchical Neutrinos}
\label{subsec:MLSM}
We have considered a specific model  for Yukawa matrices that 
gives hierarchical neutrinos. The unique feature of this model
is that it is a minimal scheme which can give TeV scale seesaw 
naturally. In this model one adds two singlets to SM; one heavy
right handed fermion, $N_R$ and one gauge singlet, $S$
having opposite lepton numbers. The most general Lagrangian 
including leptons, Higgs and extra heavy fields
can be written as 
\begin{equation}
 - {\mathcal{L}} = \bar{N_R} Y_{\nu} \tilde{\phi^{\dagger}} l_L
 + \bar{S} Y_S \tilde{\phi^{\dagger}} l_L+ \bar{S} M_R N_R^c +
 \frac{1}{2} \bar{S} \mu S^c +  \frac{1}{2} \bar{N_R} M_N N_R^c + {\rm h.c.}
 \label{eq:Gen-Lnagr}
\end{equation}
After electroweak symmetry breaking, the Higgs field $\phi$ develops
a vacuum expectation value $v/\sqrt{2}$. This results in Dirac mass
terms for the neutrinos; $m_D = Y_{\nu}\,v/\sqrt{2} $ and 
$m_S = Y_S \, v/\sqrt{2}$. The terms with coefficients $Y_S, \mu$ and
$M_N$ are lepton number violating terms. The absence of these
terms results in an enhanced symmetry of the Lagrangian. Here
we choose to work in a basis where $\mu = M_N =0$. In the basis
$(\nu_L, N_R^c, S^c)$ the neutrino mass matrix can be written as,
\begin{equation}
 \begin{pmatrix}
   0 & m_D^T & m_S^T\\
   m_D & 0 & M_R \\
  m_s & M_R^T & 0
   \end{pmatrix}
\label{eq:Numass-Lssaw}
\end{equation}
Diagonalizing the above mass matrix and assuming 
$M_R >> m_D, m_S$ the light neutrino mass matrix in the
leading order can be written as 
\begin{equation}
 m_{\rm light} = m_D^T M_R^{-1} m_S +
 m_S^T M_R^{-1} m_D
 \label{eq:lightNu-mass}
\end{equation}
The above equation for light neutrino mass matrix appears
to be linear in Dirac mass matrix ,$m_D$. Hence this particular
form of seesaw is termed as Linear seesaw. One can make an order of
magnitude calculation to see the extent of lepton number violation 
in the Lagrangian which comes from the term having coefficient $Y_S$.
Requiring $m_{\nu} = 0.1$eV and assuming $m_D = 100$ GeV, $M_R=1$ TeV
one gets $y_s = 10^{-11}$.

In light neutrino sector it can be shown that one of the
states is massless whereas the other two remain
massive. This gives us the advantage in specifying 
the massive states in terms of mass squared differences 
measured in oscillation experiments.  The Yukawa couplings
$ Y_{\nu} $ and $Y_S$ can be reconstructed from the oscillation
parameters as \cite{Gavela:2009cd}
\begin{itemize}
 \item Normal Hierarchy : $(m_1 < m_2< m_3)$
 \begin{eqnarray}
  Y_{\nu} &=& \frac{y_{\nu}}{\sqrt{2}}\, (\sqrt{1+\rho} \,U^{\dagger}_3
  + \, \sqrt{1-\rho} \, U^{\dagger}_2) \nonumber \\
  Y_S &=& \frac{y_s}{\sqrt{2}} \, (\sqrt{1+\rho} \, U^{\dagger}_3
  -\, \sqrt{1-\rho}\, U^{\dagger}_2)
  \label{eq:yukawa-NH}
 \end{eqnarray}
 
 \item Inverted Hierarchy : $(m_3 < m_2\sim m_1)$
 \begin{eqnarray}
  Y_{\nu} &=& \frac{y_{\nu}}{\sqrt{2}}\, (\sqrt{1+\rho}\, U^{\dagger}_2
  + \,\sqrt{1-\rho}\, U^{\dagger}_1) \nonumber \\
  Y_S &=& \frac{y_s}{\sqrt{2}}\, (\sqrt{1+\rho}\, U^{\dagger}_3
  -\,\sqrt{1-\rho}\, U^{\dagger}_2)
  \label{eq:yukawa-IH}
 \end{eqnarray}
 \end{itemize}
 where $y_{\nu}$ and $y_s$ are the norms of the Yukawa matrices $ Y_{\nu} $
and $Y_S$ respectively. Also 
 \begin{equation}
 \rho = \frac{\sqrt{1+r} - \sqrt{r}}{\sqrt{1+r} + \sqrt{r}}\, ({\rm NH}), \,
 \,\, \rho = \frac{\sqrt{1+r} -1}{\sqrt{1+r}+1} \, ({\rm IH})
 \end{equation}
% and 
% \begin{equation}
%  r = \frac{\Delta m^2_{\rm sol}}{\Delta m^2_{\rm atm}}
% \end{equation}

\begin{figure}[t]\centering
  \begin{tabular}{c c c}
  \includegraphics[width=0.33\textwidth,height=6.0cm]{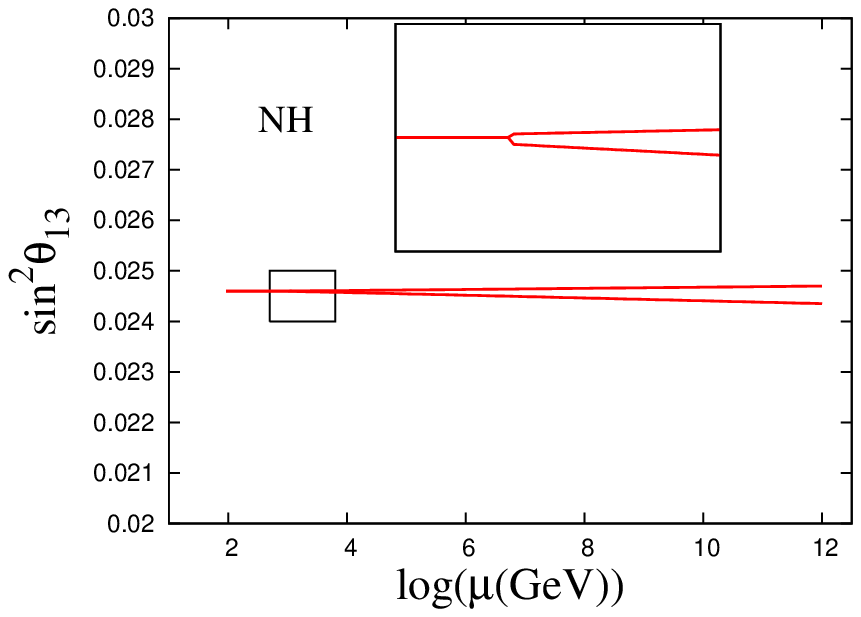} & 
  \includegraphics[width=0.33\textwidth,height=6.0cm]{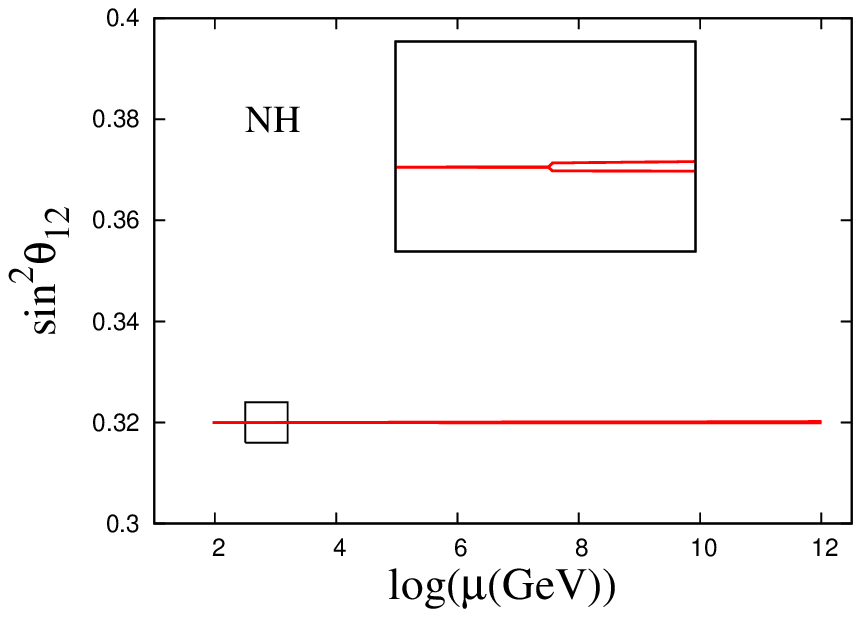} &
 \includegraphics[width=0.33\textwidth,height=6.0cm]{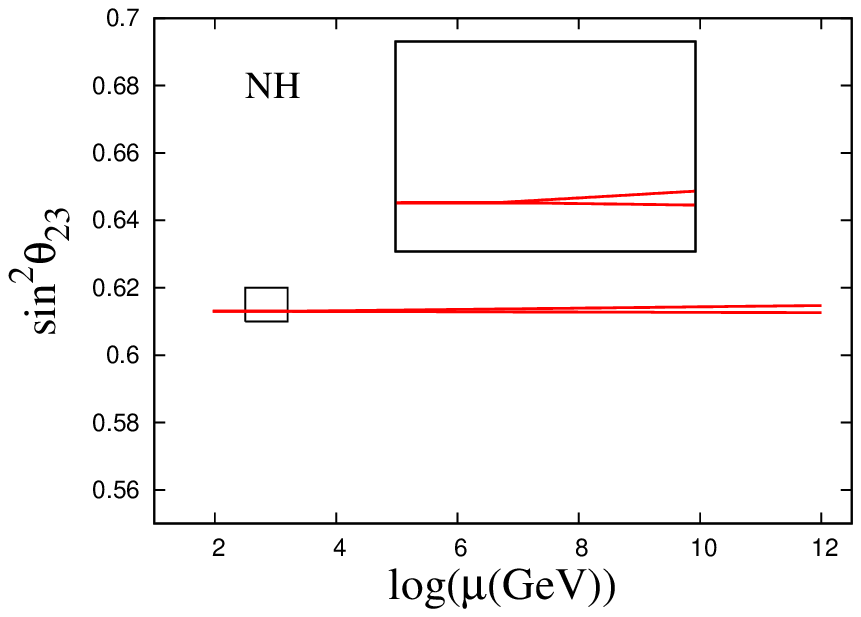}  
  \end{tabular}
\caption{\small{Running of angles for normal hierarchy from low to high scale.
At low energy we have taken the best fit values given in Table (\ref{tab:oscptm})
for normal hierarchy.
The Dirac CP phase and Majorana phases are varied randomly. The respective
figures show the maxima and mainima of the running of the angles which appear
at the threshold point as shown in the insets.}}
\label{fig:angleRun-NH}
  \end{figure}

\begin{figure}[t]\centering
  \begin{tabular}{c c c}
   \includegraphics[width=0.33\textwidth,height=6.0cm]{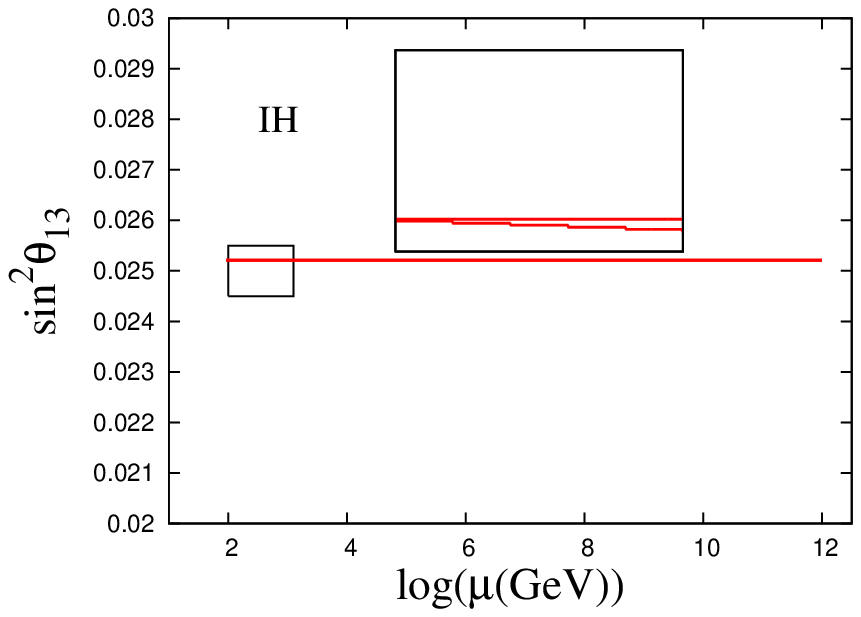} & 
  \includegraphics[width=0.33\textwidth,height=6.0cm]{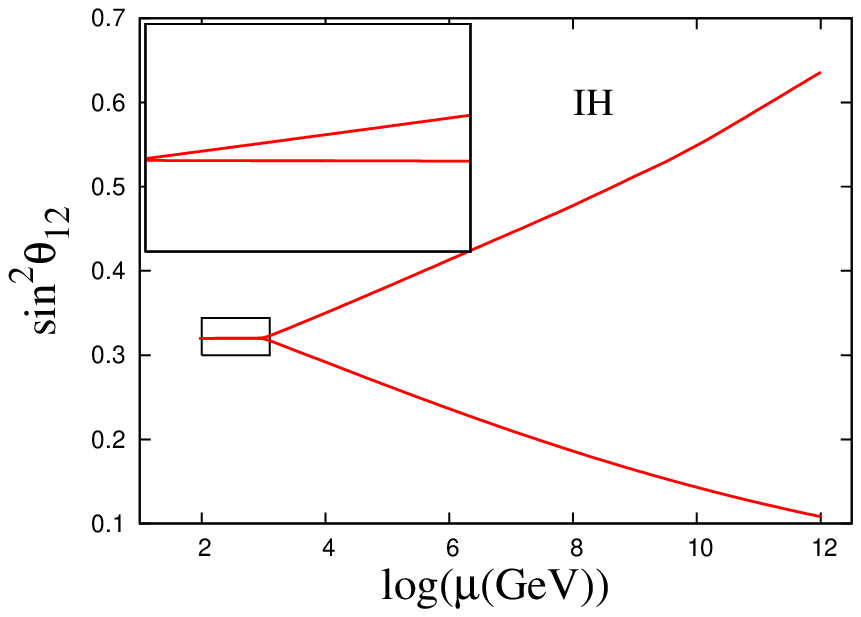} &
  \includegraphics[width=0.33\textwidth,height=6.0cm]{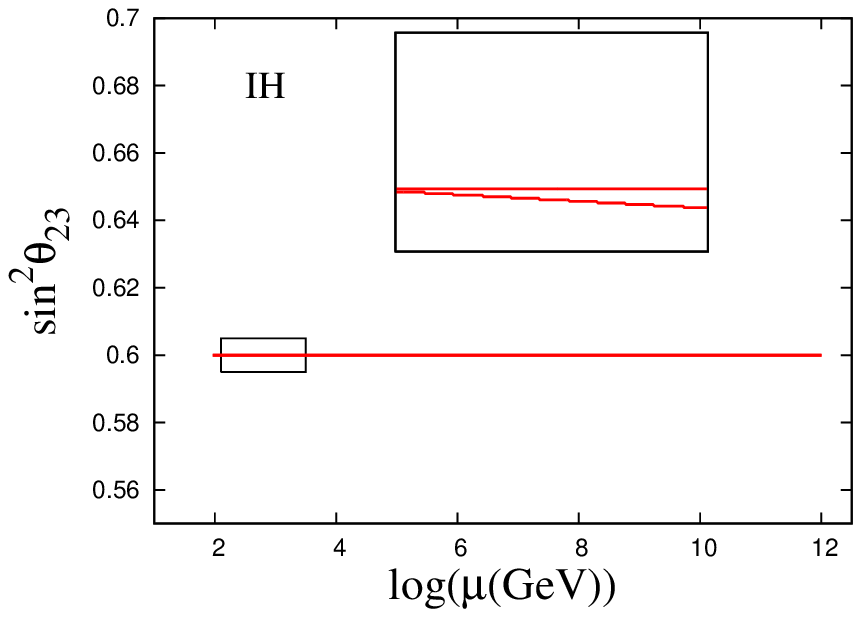}  
  \end{tabular}
\caption{\small{Running of angles for inverted hierarchy.
At low energy we have taken the best fit values given in Table (\ref{tab:oscptm})
for inverted hierarchy.
the Dirac CP phase and Majorana phases are varied randomly. The respective
figures show the maxima and minima of the running of the angles which appear
at the threshold point  as shown in the insets.}}
\label{fig:angleRun-IH}
  \end{figure}

%{\bf Running of masses}:\\
%
Figs (\ref{fig:angleRun-NH}) and (\ref{fig:angleRun-IH}) 
depict the running of the mixing angles for NH and IH respectively. 
At low scale we start with best fit values for angles given 
in Table(\ref{tab:oscptm}) for the respective hierarchies. All  
phases except leptonic phases are varied randomly. The numerical
value for $y_{\nu}$ in this case is taken to be $0.24$. 
It should be noted that in this model 
the magnitude of $y_s$ is very small, ${\mathcal{O}}(10^{-11})$. Additionally
in the running of the neutrino parameters the factor that
plays the crucial role is $Y_{\nu}^{\dagger} Y_{\nu}$. In
this case the contribution $Y_S^{\dagger} Y_S$ appears as 
an additive factor to $Y_{\nu}^{\dagger} Y_{\nu}$. Hence,
the contribution is negligible. 
In hierarchical case it appears angles do not run considerably,
except the angle, $\theta_{12}$ for IH case.
While going from low to high scale they retain their low scale values. 
The effect of threshold correction is more prominent in IH case 
than in NH case as shown in Fig.(\ref{fig:angleRun-NH})
and Fig.(\ref{fig:angleRun-IH}).
For $\theta_{13}$ since the running is proportional to $m_3$ 
this angle is not not expected to run in this model since
it has $m_3=0$. However the small amount running as seen in
the figure can be interpreted from the ${\mathcal{O}}(\sin \theta^{\Lambda}_{13})$
terms in the analytical expression.

In general it is seen that the running of angles due to RG evolution
is unidirectional i.e. either they increase or decrease while going from low to 
high scale \cite{Antusch:2003kp}. But as seen from our results the angles are running
in both directions; for example in fig.(\ref{fig:angleRun-IH}) the
angle, $\theta_{12}$ is running in both directions. This feature
comes into picture because of interplay between
different $\Delta_i$'s that appear in Eq.(\ref{eq:sin-sqr-thetas}).
The maximum running comes when the magnitude of $\Delta_e$ is dominant
as compared to $\Delta_{\mu}$ and $\Delta_{\tau}$ {\it e.g.} for a particular 
choice of phases, $\delta=0$ and $\alpha=0$, $\Delta_e \simeq 0.94$, 
$\Delta_\mu \simeq 5.7\times 10^{-4}$ and $\Delta_\tau \simeq 5.8\times 10^{-2}$. Whereas the
minimum running comes when the combined $\Delta_{\mu}$ and $\Delta_{\tau}$
are dominant in magnitude as compared to $\Delta_e$. This happens for another 
choice of phases, $\delta=0$ and $\alpha=\pi/2$ resulting $\Delta_e \simeq 3.1\times 10^{-2}$, 
$\Delta_\mu \simeq 0.42$ and $\Delta_\tau \simeq 0.55$. Thus this
feature is unique to RG evolution including threshold effect.
Also from Eq.(\ref{eq:kij-IH}) it is easy to check that the
running of $\theta_{12}$ is proportional to $1/r$ which 
enhances the effect. 

\begin{figure}
%\FIGURE{
%\hspace{-0.5cm}
\begin{center}
\includegraphics[width=9.0cm,height=8.0cm]{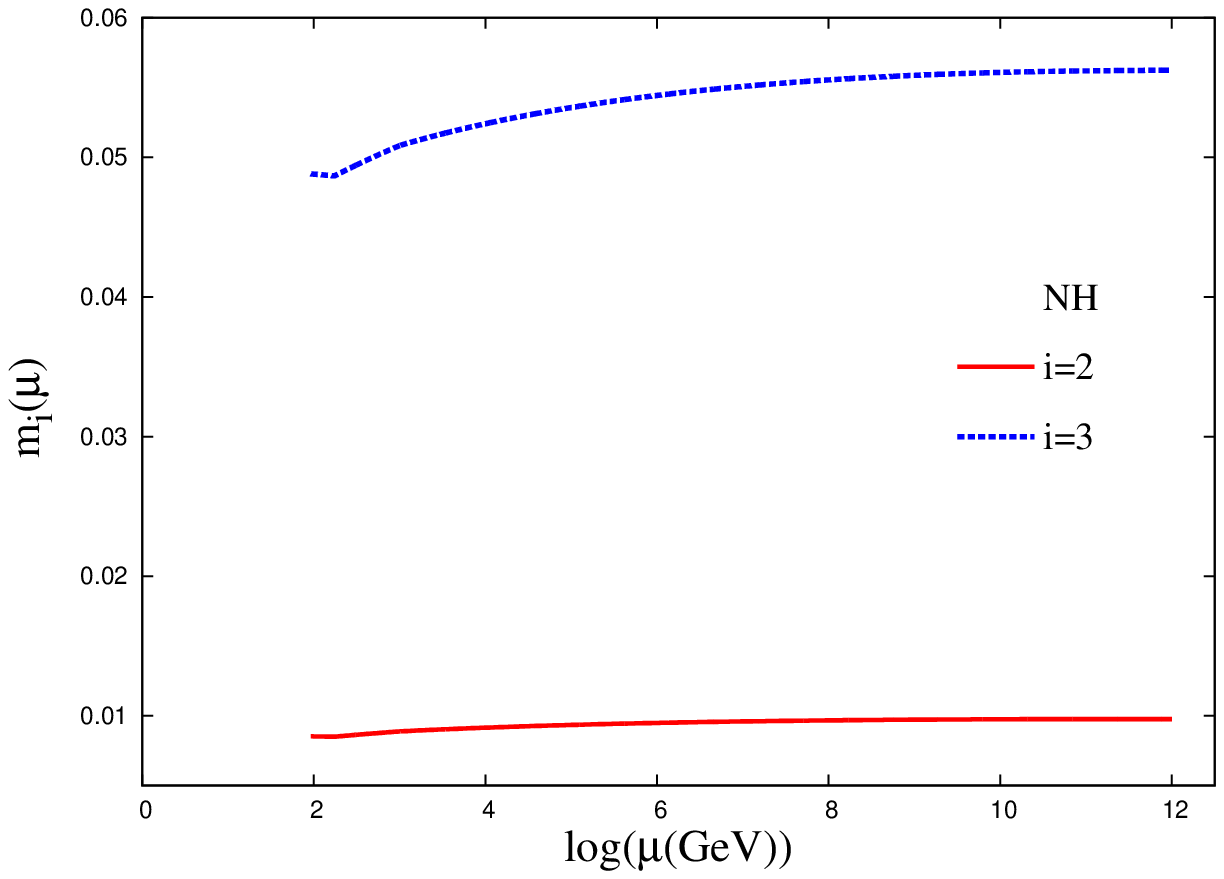}
\end{center}
\caption{\small{Running masses, $m_2(\mu)$ and $m_3(\mu)$
for normal hierarchy from low scale to high scale. For this case
we have taken $m_1 =0$ and the other two masses are reconstructed from
experimental values at low energy.}}
\label{fig:RunningMasses-NH}%}
\end{figure}
\begin{figure}
%\FIGURE{
%\hspace{-0.5cm}
\begin{center}
\includegraphics[width=9.0cm,height=8.0cm]{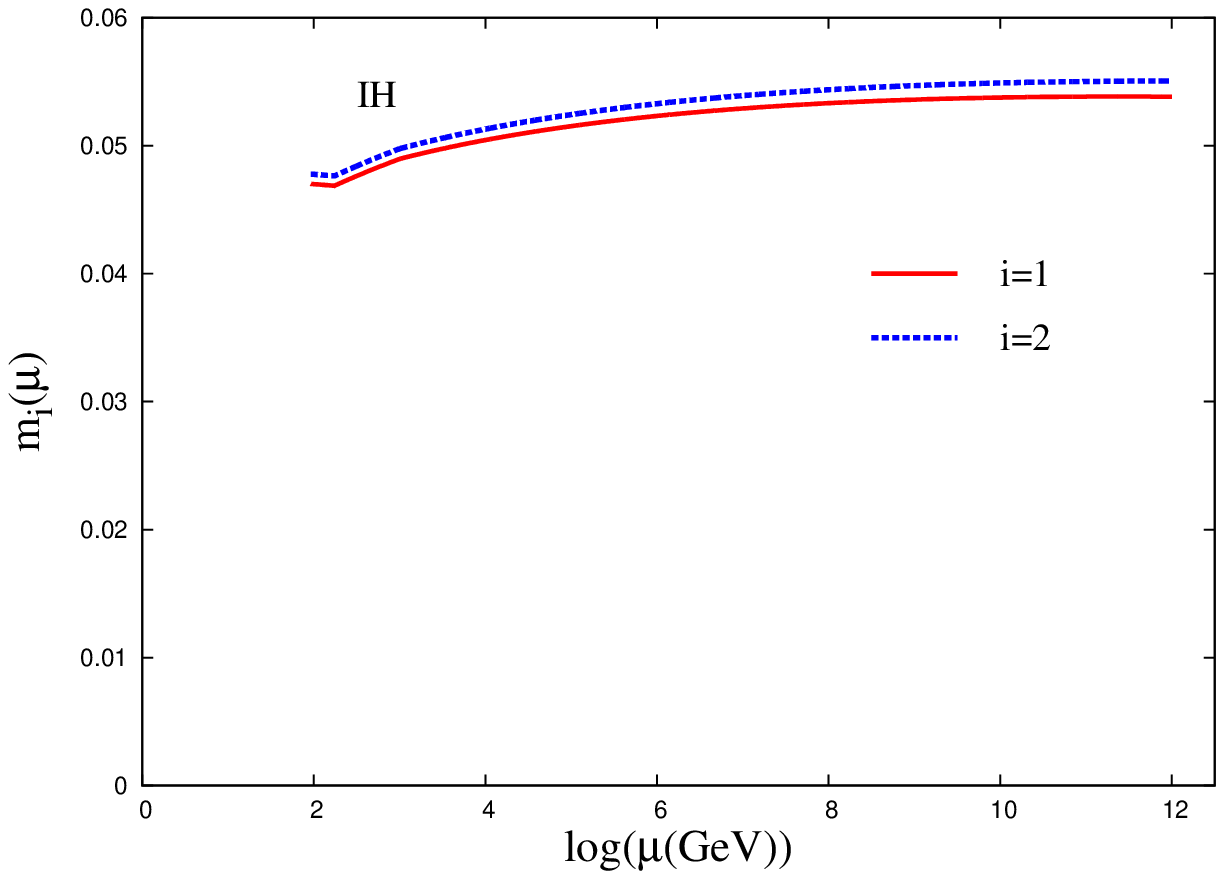}  
\end{center}
\caption{\small{Running of masses $m_1(\mu)$ and $m_2(\mu)$ for 
inverted hierarchy from low scale to high scale.For this case
we have taken $m_3 =0$ and the other two masses are reconstructed from
experimental values at low energy.}}
\label{fig:RunningMasses-IH}%}
  \end{figure}
\begin{figure}
%\FIGURE{
%\hspace{-0.5cm}
\begin{center}
\includegraphics[width=9.0cm,height=8.0cm]{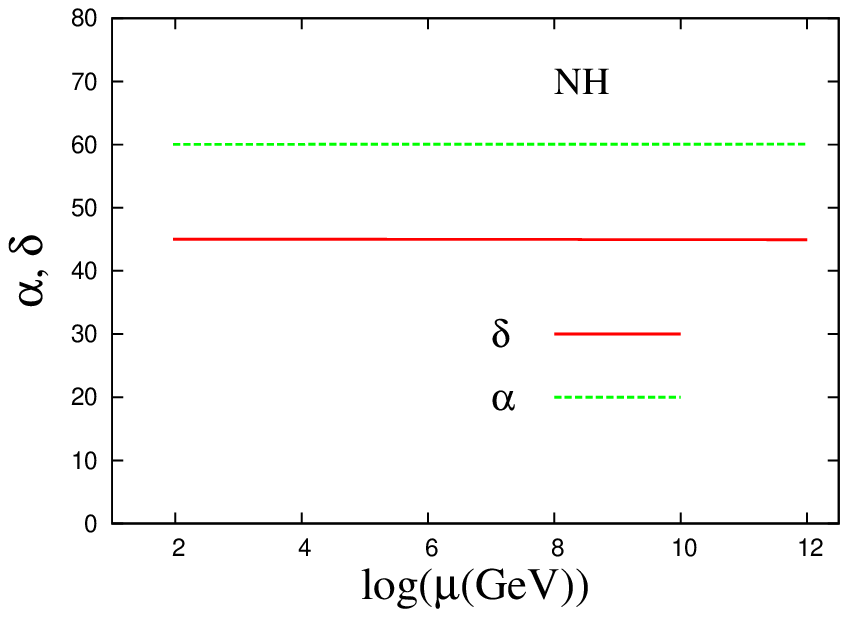}  
\end{center}
\caption{\small{Running phases
for normal hierarchy from low scale to high scale. Since in this case
one of the masses, $m_1 =0$ there is one independent Majorana phase.}}
\label{fig:phasesNH}%}
\end{figure}
\begin{figure}
%\FIGURE{
%\hspace{-0.5cm}
\begin{center}
\includegraphics[width=9.0cm,height=8.0cm]{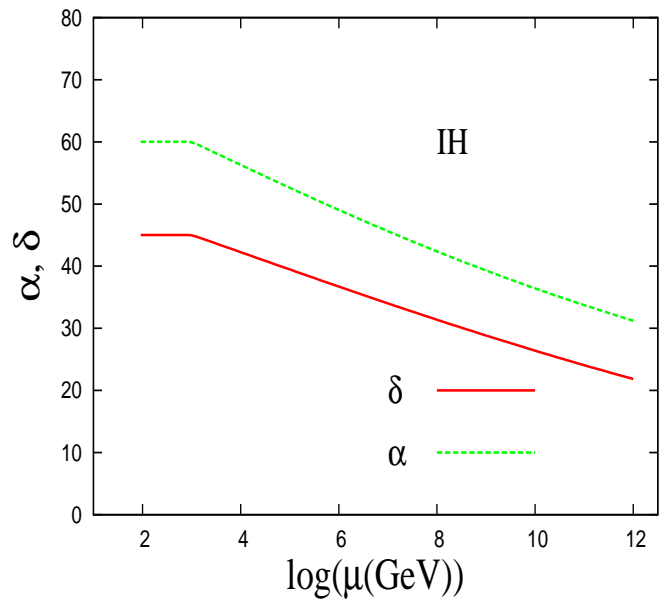} 
\end{center}
\caption{\small{Running phases
for inverted hierarchy from low scale to high scale. In this case
one of the masses, $m_3 =0$ and hence there is one independent Majorana phase.}}
\label{fig:phasesIH}%}
\end{figure}

The Figs.(\ref{fig:RunningMasses-NH}) and (\ref{fig:RunningMasses-IH})
show the running of the masses. It appears that the masses
do not run much. Form the analytical expressions for
masses given in Eq.(\ref{eq:mass-exp}), one can see the 
running of masses is proportional to the respective masses
themselves at leading order. For NH the masses, $m_1$ and
$m_2$ are plotted with renormalization scale as $m_3 = 0$
in this case. Since for IH case $m_3 =0$, the masses
$m_1$ and $m_2$ are plotted. 
In Figs. \ref{fig:phasesNH} and \ref{fig:phasesIH} 
we show the running of the phases. In this case since one 
of the mass eigenvalues is zero there is only one 
independent Majorana phase. 
The figures demonstrate that for NH
the phases do not run.
This feature can be understood from the analytical 
expressions given in Eq.(\ref{eq:a-i-eqns}).
One can see the running of the Majorana phases 
is proportional to $m_1$ which is vanishing for normal 
hierarchy in this model. 
However for IH there is considerable running of the phases. 
For inverted hierarchy the phases run considerably because
of the enhancement coming from the solar mass squared
difference that appears in the denominator of $a_i$'s in
Eq.(\ref{eq:a-i-eqns}). 

\subsection{Quasidegenrate Neutrinos}
\label{subsec:Qsi-deg}
%One often adds three right handed heavy neutrinos, $N_R$ to the theory
%to get nonzero neutrino masses. In this case the Lagrangian is
%given by,
%\begin{equation} 
%-{\cal{L}} = (Y_{\nu}) \bar{N_R} \tilde{\phi}^\dagger l_L +
%Y_l {\bar{E}_R} \phi l_L + \frac{1}{2}
%{\bar{N_R}^c} {\bf{M_R}} N_R + h.c.
%\label{eq:yyukawa-QD} 
%\end{equation} 
%where $\tilde{\phi} = i \sigma^2 \phi^{*}$ , and 
%$Y_{\nu}$ is the neutrino Dirac Yukawa coupling matrix. $M_R$ is mass
%scale below which all the heavy neutrinos decouple from the theory
%to give an effective mass matrix for the neutrinos which is given by
%\begin{equation}
% m_{\nu} = \frac{v^2}{2} Y_{\nu}^T M_R^{-1} Y_{\nu}
%\end{equation}
%where $v$ is the vev of the Higgs doublet, $H$. 
In this section we consider adding three heavy right handed
neutrinos, $N_R$ degenerate in mass to the SM. The Lagrangian
is given in Eq.(\ref{eq:yyukawa}).Here $Y_{\nu}$ is a $3 \times 3$
matrix.Using the parametrization
in \cite{Casas:2001sr} $Y_{\nu}$ can be reconstructed from Eq.~(\ref{seesaw1})
as
\begin{equation}
 Y_{\nu} = \frac{2}{v^2} \sqrt{M^d_R} R \sqrt{m^d_{\nu}} U^{\dagger}
{\label{eq:Ynu-parmtz}}
\end{equation}
where $M^d_R$ and $m^d_{\nu}$ are the diagonalized mass matrix for
heavy and light neutrinos respectively. $U$ is the standard $U_{\rm PMNS}$ 
matrix diagonalizing the mass matrix $U m_{\nu} U^{\dagger} = m^d_{\nu}$.
$R$ is a complex orthogonal matrix, $R R^T = I$. The $R$ matrix can be
parametrized  as
\begin{equation}
 R = O e^{i A}
\end{equation}
where $O$ and $A$ are real matrices. The condition of orthogonality
implies that $O$ is orthogonal and $A$ is antisymmetric. 
\begin{equation}
 \begin{pmatrix}
 0 & a & b\\
 -a & 0 & c \\
-b & -c & 0 
 \end{pmatrix}
 \label{eq:Amatrix}
\end{equation}
with real $a,b,c$. In particular
\begin{equation}
 e^{i A} = I - \frac{\cosh \omega -1}{\omega^2} A^2 + i \frac{\sinh \omega}{\omega} A
 \label{eq:expA}
\end{equation}
where $\omega=\sqrt{a^2 +b^2 + c^2}$. Using Eq.(\ref{eq:Amatrix}) and Eq.(\ref{eq:expA})
in Eq.(\ref{eq:Ynu-parmtz}) we can get $Y_{\nu}$ for our numerical work. It can be 
observed that the parameter $\omega$ plays an important role in determining the magnitude
of $Y_{\nu}$. For our numerical work we have taken $\omega = 11.6$ and $a=b=c$. Similar kind of
assumption can be found in \cite{Rodejohann:2012px, Pascoli:2003rq}. For this particular
choice of $\omega$ and with $M_R$ at $1$ TeV, the order of 
 ${\rm Tr}(Y^{\dagger}_{\nu} Y_{\nu}) \simeq 0.06$.

\begin{figure}[t]\centering
%  \begin{tabular}{c c c}
%   \includegraphics[width=9.0cm,height=8.0cm]{angle-vs-mu.eps}
\includegraphics[width=9.0cm,height=8.0cm]{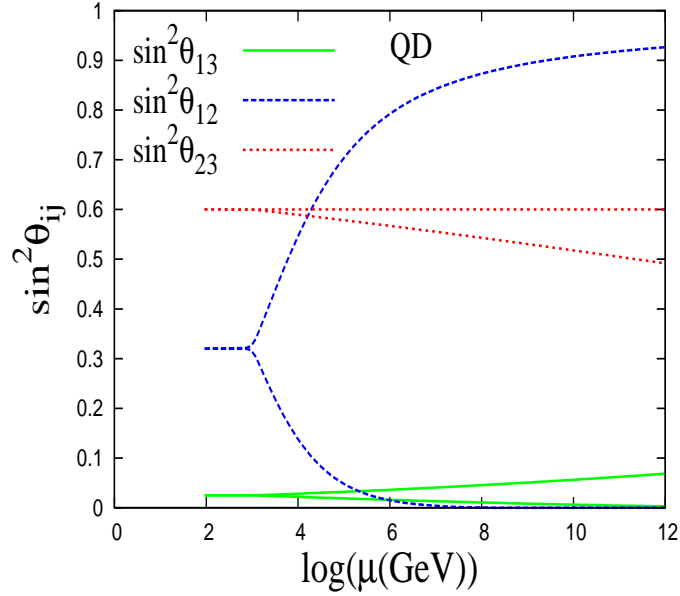} 
%\includegraphics[width=9.0cm,height=8.0cm]{angle-vs-mu.pdf} 
% \epsfig{file=theta12-deg.eps,width=0.33\textwidth,height=6.0cm} &
% \epsfig{file=theta23-deg.eps,width=0.33\textwidth,height=6.0cm}  
%  \end{tabular}
\caption{\small{Running of angles for quasidegenerate neutrinos.
At low energy we have taken the common mass $m_0 = 0.2$eV.
The Dirac CP phase and Majorana phases are varied randomly. The respective
figures show the maxima and minima of the running of the angles which appear
at the threshold point.}}
\label{fig:angleRun-deg}
  \end{figure}

The Fig. (\ref{fig:angleRun-deg})  shows the running of the mixing angles 
for QD neutrinos. The angles seem to run considerably covering
a wide range of values at high scale. In particular at high scale $\sin^2 \theta_{12}$
can accommodate the range from zero to $\simeq 0.92$.
Both the angles; $\theta_{12}$ and $\theta_{13}$
exhibit bidirectional running whereas the running of $\theta_{23}$ is 
unidirectional. This behavior of running of angles can be understood from
 the analytical expressions given in Eqs.(\ref{eq:sin-sqr-thetas})
and (\ref{eq:kij-QD}) and with specific numerical values of $\Delta_e \,, \Delta_\mu \,\text{and}\, \Delta_\tau$. 
For example we find for a particular choice of phase, the magnitude 
of $\Delta_e \simeq 3.08\times 10^{-2}$, $\Delta_{\mu} \simeq 1.9 \times 10^{-2}$ and
$\Delta_{\tau} \simeq 6.13 \times 10^{-3}$. For some other choice of phases, 
the magnitude of $\Delta_e \simeq 3.3 \times 10^{-3}$, $\Delta_{\mu} \simeq 5.5 \times 10^{-2}$ and
$\Delta_{\tau} \simeq 4.8 \times 10^{-3}$. Also in our numerical study, %shows that the magnitude of
$\Delta_{\mu}$ is always greater than that of $\Delta_{\tau}$ whereas 
the magnitude of $\Delta_e$ can be more or less than that of $\Delta_{\mu}$.
For the case of angle $\theta_{13}$, though
$\Delta_{\mu}$ is always greater than $\Delta_{\tau}$ the relative sign difference
between $\cos \alpha_1$ and $\cos \alpha_2$ appearing in the expression of 
$K_{\mu 13}$ results in bidirectional running of the same. So even if at high scale
the angle $\theta_{13}$ can be of zero value the threshold effect can result in nonzero
value at low scale. 

\begin{figure}
%\FIGURE{
%\hspace{-0.5cm}
\begin{center}
\includegraphics[width=9.0cm,height=8.0cm]{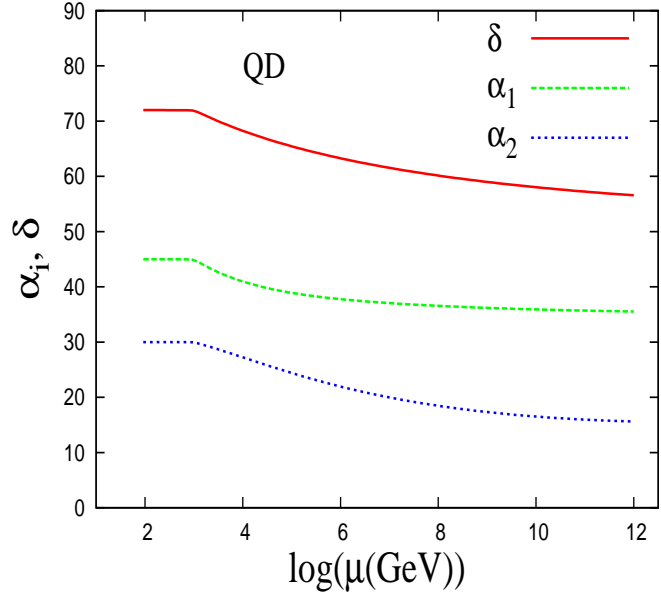}  
\end{center}
\caption{\small{Running of phases
for quasidegenerate neutrinos from low scale to high scale.}}
\label{fig:PhasesDeg}%}
\end{figure}

In general the running for the QD case 
is more than that of hierarchical case, in presence of threshold effects.
This is highly dependent on the magnitude and form of $Y_\nu$.
And using Casas-Ibarra parametrization the magnitude of $Y_{\nu}$ 
can be magnified even for a low seesaw scale.
Fig.(\ref{fig:PhasesDeg}) shows the running of the phases. 
One can see the effect of threshold correction on the running 
of phases at $1$ TeV. This also depends on the form and magnitude of $Y_\nu$ chosen.

% \sout {In this case only the mixing angle $\theta_{12}$ run 
% considerably and depending on the phases one can reach 
% $\sin^2 \theta_{12} \simeq 0.42$ at high scale. 
% One can see comparing the running of the mixing angle $\theta_{12}$ 
% that the running of $\theta_{12}$ is higher for the IH case of the 
% minimal linear seesaw model considered previously. 
% This is because in this scenario the magnitude of the 
% Yukawa coupling remains small in order to ensure the mass
% of the light neutrinos small with the scale of  $M_R$  to be
% $O(TeV)$. Thus the running is less than the linear seesaw model 
% considered. Thus although in general we expect that the running for the QD case 
% is more than that of hierarchical case, in presence of threshold effects
% this is highly dependent on the magnitude and form of $Y_\nu$. 
% Fig.(\ref{fig:PhasesDeg}) shows the running of the phases. 
% }

% 
  
\section{Conclusion}
\label{sec:concl}
We consider the threshold effect in renormalization group (RG) 
evolution of the neutrino masses and mixing arising in 
TeV scale seesaw models. In such models the heavy states stay 
coupled to the theory once their mass threshold is crossed 
and can give rise to additional terms in the beta functions. 
We obtain the analytical expressions for the coefficients
governing the running of masses, mixing angles and phases
using the factorization technique including these additional terms. 
The threshold corrections can give rise to an enhanced running
effect with the coefficient $\Delta_{\tau} \simeq 10^{-3}$ as compared 
to $\Delta_{\tau} \simeq 10^{-5}$ where corrections due to
threshold effect are absent. The actual running depends on the 
form of the Yukawa coupling matrix $Y_\nu$.  
We perform a numerical analysis of this in the bottom up approach 
using two specific models where the matrix $Y_\nu$ can be reconstructed 
easily. The first case considered by us is the minimal linear seesaw model
with hierarchical light neutrinos having one of the mass eigenvalues as zero. 
The heavy neutrinos are pseudo-Dirac in nature and both have mass 
$\sim$ TeV. Thus the model contains a single threshold at the TeV scale. 
In this case for NH we do not obtain appreciable running for 
the mixing angles, masses and phases. For IH, the mixing angle $\theta_{12}$ 
and the phases show considerable running effect. The unique feature
of threshold effect is that the mixing angle $\theta_{12}$ can increase
or decrease from low to high scale depending on initial choice of 
phases i.e the running is not unidirectional. The second example that we consider 
is a model with quasidegenerate light neutrinos. For this case we reconstruct the 
Yukawa matrix in terms of the low scale parameters using the 
Casas-Ibarra parametrization where the three heavy neutrinos 
are taken to be degenerate in mass. For this case we obtain 
considerable running effect for the  mixing angle and the phases.
In this case two out of the three mixing angles ($\theta_{12}\,,\theta_{13}$)
run in both directions. Also the running effect is more than that of minimal
linear seesaw model. 
To conclude, in presence of threshold effects the running of the 
neutrino parameters depend on the form of the Yukawa matrix apart 
from the mass spectrum and the phases. 
\begin{center}
{\bf Acknowledgment}
\end{center}
The authors would like to thank Amol Dighe for useful discussions.

%For our numerical study we need a definite form of neutrino Yukawa
%matrix, $y_{\nu}$. We have considered two cases for reconstructing
%the $Y_{\nu}$ and perform the RG evolution from low to high scale.
%In the first case which is the case of hierarchical neutrinos our
% $Y_{\nu}$ is reconstructed form low energy oscillation parameters.
%In the second case, i.e for quasidegenerate neutrinos the $Y_{\nu}$
%is parametrized by Casas-Ibarra formalism.

%{\bf The symbol $\lambda$ has been used twice; one for low scale other for 
%higgs self coupling. Also $r$, in degenerate case and non-deg case}

\appendix
\section{Analytical results for Majorana and leptonic phases}
\label{sec:app}
In this section we give the analytic expressions of the coefficients
that appear in the perturbation of Dirac, Majorana and leptonic phases
(the analytic expressions of the coefficients of the mixing angles for
non zero $\theta_{13}$ can be found in \cite{Bergstrom:2010qb}).
\begin{eqnarray}
\delta^{\lambda} &=& \delta^{\Lambda} + d_e \Delta_e+
d_{\mu} \Delta_{\mu} +  d_{\tau}\Delta_{\tau}  \nonumber \\
\alpha^{\lambda}_i &=& \alpha^{\Lambda}_i+  a_{e_i}\Delta_e +
 a_{\mu_i} \Delta_{\mu} +  a_{\tau_i}\Delta_{\tau} 
%\label{eq:majorana-Dirac-low-high}
\end{eqnarray}
%
%The leptonic phases at high scale to that at low scale can be
%related as,
%
\begin{equation}
  \Phi_i^{\lambda} = \Phi_i^{\Lambda}+  p_{e_i}\Delta_e +
 p_{\mu_i} \Delta_{\mu} + p_{\tau_i}\Delta_{\tau} 
% \label{eq:leptonic-low-high}
\end{equation}
We define a quantity $\xi$ such that
\begin{equation}
 \xi_{ij} = \frac{m_i - m_j}{m_i + m_j}
\end{equation}
The expressions are given by
\begin{eqnarray}
\label{de}
d_e &=& -\,\frac{1}{2}\left( \frac{1}{\xi_{12}}- \xi_{12} \right) \sin\left(\alpha_1^{\Lambda} - \alpha_2^{\Lambda} \right) + \frac{1}{2}\left[\left( \frac{1}{\xi_{13}}- \xi_{13} \right) \cos^2\theta_{12}^{\Lambda} \sin\left(\alpha_1^{\Lambda} - 2\delta^{\Lambda} \right) \right . \nonumber \\ 
& & - \,\left( \frac{1}{\xi_{32}}- \xi_{32} \right) \sin^2\theta_{12}^{\Lambda} \sin\left(\alpha_2^{\Lambda} - 2\,\delta^{\Lambda} \right) \bigg] \nonumber \\
& & + \,2 \cot 2\,\theta_{23}^{\Lambda} \left[\frac{1}{4}\left( \frac{1}{\xi_{13}}- \xi_{13} \right)\sin\left(\alpha_1^{\Lambda} - \delta^{\Lambda} \right) + \frac{1}{4}\left( \frac{1}{\xi_{32}}- \xi_{32} \right) \sin\left(\alpha_2^{\Lambda} - \delta^{\Lambda} \right) \right . \nonumber \\ 
& & \left . + \,\frac{1}{4} \left(\frac{m_3^{\Lambda}}{m_2^{\Lambda}} \left( \frac{1}{\xi_{32}}- \xi_{32} \right) + \frac{m_3^{\Lambda}}{m_1^{\Lambda}} \left( \frac{1}{\xi_{13}}- \xi_{13} \right) \right) \sin\delta^{\Lambda} \right] \sin 2\theta_{12}^{\Lambda} \sin\theta_{13}^{\Lambda} \nonumber \\
& & - \,\text{cosec}\,\delta^{\Lambda} \text{cosec}\,\theta_{12}^{\Lambda} \,\text{cosec}\,\theta_{23}^{\Lambda} \,\sec\theta_{12}^{\Lambda}\, \sec\theta_{23}^{\Lambda} \left[\frac{1}{2}\left\{ -\frac{1}{2}\left( \frac{1}{\xi_{13}}- \xi_{13} \right) \times \right . \right . \nonumber \\ 
& & \left . \left . \cos\left(\alpha_1^{\Lambda} - \delta^{\Lambda} \right) + \left(-\frac{1}{2} \left(\frac{1}{\xi_ {32}}-\xi_{32}  \right) \cos\left(\alpha_2^{\Lambda} - \delta^{\Lambda} \right) + \frac{1}{4}\left( \frac{1}{\xi_{12}}- \xi_{12} \right) \right . \right . \right . \nonumber \\
& &\left . \left . \left( \left( \frac{m_1^{\Lambda}}{m_2^{\Lambda}} + \frac{m_2^{\Lambda}}{m_1^{\Lambda}} \right) \cos\delta^{\Lambda} + 2 \cos\left(\alpha_1^{\Lambda} - \alpha_2^{\Lambda} + \delta^{\Lambda} \right) \right) \right) \right\} \cos \theta_{12}^{\Lambda} \sin^3 \theta_{12}^{\Lambda} \nonumber \\
& &+ \,2 \cos\delta^{\Lambda}\left\{ \left( - 1 - \frac{1}{4}\frac{m_1^{\Lambda}}{m_3^{\Lambda}}\left(\frac{1}{\xi_{13}}-  \xi_{13} \right) - \frac{3}{4}\frac{m_2^{\Lambda}}{m_3^{\Lambda}} \left( \frac{1}{\xi_{32}}- \xi_{32} \right) \right) \right . \nonumber \\
& & + \,\left(\frac{1}{2}\left( \left( \frac{m_1^{\Lambda}}{m_ 2^{\Lambda}} + \frac{m_2^{\Lambda}}{m_1^{\Lambda}} \right) \left(  \frac{1}{\xi_ {12}}- \xi_{12} \right) - \frac{1}{2}\frac{m_2^{\Lambda}}{m_3^{\Lambda}} \left( \frac{1}{\xi_{32}} - \xi_{32}  \right) - \frac{1}{2}\frac{m_1^{\Lambda}}{m_3^{\Lambda}}\left( \frac{1}{\xi_{13}}- \xi_{13} \right) \right) \right .  \nonumber
\end{eqnarray}
\begin{eqnarray}
& & \left . + \,\left( \frac{1}{\xi_{12}}- \xi_{12} \right) \cos\left(\alpha_1^{\Lambda} - \alpha_2^{\Lambda} \right) \right) \cos 2\theta_{12}^{\Lambda} - \frac{1}{4} \left( \frac{1}{\xi_ {13}}- \xi_{13} \right) \cos\left(\alpha_1^{\Lambda} - 2 \delta^{\Lambda} \right) \times \nonumber \\
& & \left . \left(- 1 + 3 \cos 2\theta_{12}^{\Lambda} \right) - \frac{1}{4} \left( \frac{1}{\xi_{32}} - \xi_{32}  \right) \cos\left(\alpha_2^{\Lambda} - 2\delta^{\Lambda} \right) \left(1 + 3 \cos 2\theta_{12}^{\Lambda} \right) \right\} \times \nonumber \\
& & \sin 2\theta_{12}^{\Lambda} \bigg] \sin^2\theta_{13}^{\Lambda} \sin 2\theta_{23}^{\Lambda}
\end{eqnarray}
%\newpage
\begin{eqnarray}
\label{dmu}
d_\mu &=& \frac{1}{8}\left[\left( \frac{1}{\xi_{13}}- \xi_{13} \right)\left(\frac{m_3^{\Lambda}}{m_1^{\Lambda}}\sin\delta^{\Lambda} - \sin\left(\alpha_1^{\Lambda} - \delta^{\Lambda} \right) \right) + \left( \frac{1}{\xi_{32}}- \xi_{32} \right)\left(\frac{m_3^{\Lambda}}{m_2^{\Lambda}}\,\sin\delta^{\Lambda} \right . \right . \nonumber \\
& & - \,\sin\left(\alpha_2^{\Lambda} - \delta^{\Lambda} \right) \bigg) \bigg] \frac{\sin 2\theta_{12}^{\Lambda} \sin 2\theta_{23}^{\Lambda}}{\sin \theta_{13}^{\Lambda}}\nonumber \\
&+& \frac{1}{2}\left[\left( \frac{1}{\xi_{12}}- \xi_{12} \right)\cos^2\theta_{23}^{\Lambda}\sin\left(\alpha_1^{\Lambda} - \alpha_2^{\Lambda} \right) + \left( \frac{1}{\xi_{32}}- \xi_{32} \right) \Big(\cos^2\theta_{12}^{\Lambda} \cos 2\theta_{23}^{\Lambda} \sin \alpha_2^{\Lambda} \right . \nonumber \\
& & + \,\sin^2\theta_{12}^{\Lambda} \sin^2\theta_{23}^{\Lambda} \sin\left(\alpha_2^{\Lambda} - 2\delta^{\Lambda} \right) \Big) - \left( \frac{1}{\xi_{13}}- \xi_{13} \right)\Big(\sin^2\theta_{12}^{\Lambda} \cos 2\theta_{23}^{\Lambda} \sin \alpha_1^{\Lambda} \nonumber \\
& & + \,\cos^2\theta_{12}^{\Lambda} \sin^2\theta_{23}^{\Lambda} \sin\left(\alpha_1^{\Lambda} - 2\delta^{\Lambda} \right)\Big) \bigg]\nonumber \\
&+& \left[\text{cosec}2\theta_{12}^{\Lambda} \left\{\cos^2\theta_{12}^{\Lambda}\left(\frac{1}{2}\left( \frac{1}{\xi_{12}}- \xi_{12} \right)\sin\left(\alpha_1^{\Lambda} - \alpha_2^{\Lambda} - \delta^{\Lambda} \right) + \frac{1}{4}\left( \frac{1}{\xi_{32}}- \xi_{32} \right) \times \right . \right . \right . \nonumber \\
& & \left(- 1 + 3 \cos 2\theta_{12}^{\Lambda} \right)\sin\left(\alpha_2^{\Lambda} - \delta^{\Lambda} \right) \bigg) + \left(\frac{1}{16}\left(4\left(\frac{m_1^{\Lambda}}{m_2^{\Lambda}} + \frac{m_2^{\Lambda}}{m_1^{\Lambda}}\right) \left(\xi_{12} - \frac{1}{\xi_{12}}\right) \right . \right . \nonumber \\
& & \left . - \,5 \,\frac{m_2^{\Lambda}}{m_3^{\Lambda}}\left( \frac{1}{\xi_{32}}-\xi_{32} \right) - 5\,\frac{m_1^{\Lambda}}{m_3^{\Lambda}}\left( \frac{1}{\xi_{13}}-\xi_{13}\right) \right) + \left(\frac{1}{4}\left(\frac{m_1^{\Lambda}}{m_3^{\Lambda}}\left( \frac{1}{\xi_{13}}-\xi_{13}\right) \right . \right . \nonumber \\
& & \left . - \,\frac{m_3^{\Lambda}}{m_2^{\Lambda}}\left(\frac{1}{\xi_{32}}-\xi_{32} \right)\right)\cos 2\theta_{12}^{\Lambda} + \frac{1}{16}\left(\frac{m_3^{\Lambda}}{m_2^{\Lambda}}\left(\frac{1}{\xi_{32}}-\xi_{32} \right) + \frac{m_3^{\Lambda}}{m_1^{\Lambda}}\left( \frac{1}{\xi_{13}}-\xi_{13} \right)\right)\times \nonumber \\
& & \cos 4\theta_{12}^{\Lambda} \bigg) \bigg)\sin\,\delta^{\Lambda} - \left(\frac{1}{4}\left(\frac{1}{\xi_{13}}-\xi_{13} \right)\left(1 + 3 \cos 2\theta_{12}^{\Lambda} \right)\sin\left(\alpha_1^{\Lambda} -  \delta^{\Lambda} \right) \right . \nonumber \\
& & \left . + \,\frac{1}{2}\left(\frac{1}{\xi_{12}}-\xi_{12}\right)\sin\left(\alpha_1^{\Lambda} - \alpha_2^{\Lambda} + \delta^{\Lambda} \right) \right)\sin^2\theta_{12}^{\Lambda} \bigg\}\sin 2\theta_{23}^{\Lambda} + \left\{\frac{1}{4}\left(\frac{1}{\xi_{13}}-\xi_{13} \right) \times \right . \nonumber \\
& & \sin\left(\alpha_1^{\Lambda} - \delta^{\Lambda} \right) + \frac{1}{4}\left(\frac{1}{\xi_{32}}-\xi_{32} \right)\sin\left(\alpha_2^{\Lambda} - \delta^{\Lambda} \right) + \frac{1}{4}\left(\frac{m_3^{\Lambda}}{m_2^{\Lambda}}\left( \frac{1}{\xi_{32}}-\xi_{32} \right) \right . \nonumber \\
& & \left . \left . + \,\frac{m_3^{\Lambda}}{m_1^{\Lambda}}\left( \frac{1}{\xi_{13}}-\xi_{13}\right) \right)\sin \delta^{\Lambda} \right\}\sin 2\theta_{12}^{\Lambda} \tan \theta_{23}^{\Lambda} \bigg]\sin \theta_{13}^{\Lambda}\nonumber \\
% \end{eqnarray}
% \begin{eqnarray}
&+& \left[-\frac{3}{16}\left( \frac{1}{\xi_{13}}-\xi_{13} \right)\cos\left(\alpha_1^{\Lambda} - \delta^{\Lambda} \right) - \frac{5}{16}\left( \frac{1}{\xi_{32}}-\xi_{32} \right)\cos\left(\alpha_2^{\Lambda} - \delta^{\Lambda} \right) \right . \nonumber \\
& & + \,\frac{1}{8}\left\{-\frac{m_1^{\Lambda}}{m_3^{\Lambda}}\left( \frac{1}{\xi_{13}}-\xi_{13}\right) + 2 \frac{m_2^{\Lambda}}{m_1^{\Lambda}}\left( \frac{1}{\xi_{12}}-\xi_{12}\right) - 3 \frac{m_2^{\Lambda}}{m_3^{\Lambda}}\left( \frac{1}{\xi_{32}}-\xi_{32}\right) \right\}\cos\,\delta^{\Lambda} \nonumber\\
& & + \, \frac{1}{4}\left(\frac{1}{\xi_{32}}-\xi_{32} \right)\cos\left(\alpha_1^{\Lambda} - \alpha_2^{\Lambda} + \delta^{\Lambda} \right) - \frac{1}{16}\left( \frac{1}{\xi_{13}}-\xi_{13} \right)\cos\left(\alpha_1^{\Lambda} - 3\delta^{\Lambda} \right)\times \nonumber
\end{eqnarray}
\begin{eqnarray}
& & \left(- 1 + 3 \cos 2\theta_{12}^{\Lambda} \right) - \frac{1}{16}\left( \frac{1}{\xi_{32}}-\xi_{32} \right)\cos\left(\alpha_2^{\Lambda} - 3\delta^{\Lambda} \right)\left(1 + 3 \cos 2\theta_{12}^{\Lambda} \right) \nonumber \\
& & + \, \cos 2\theta_{12}^{\Lambda} \left\{\left(\frac{1}{16}\left(\frac{1}{\xi_{13}}-\xi_{13} \right) + \frac{1}{4}\left( \frac{1}{\xi_{32}}-\xi_{32}\right)\cos \alpha_2^{\Lambda}\right)\cos\left(\alpha_1^{\Lambda} - \delta^{\Lambda} \right) \right . \nonumber \\
& & + \frac{1}{16}\left(\frac{1}{\xi_{32}}-\xi_{32} \right)\cos\left(\alpha_2^{\Lambda} - \delta^{\Lambda} \right) + \frac{1}{8}\left(\left(\frac{m_1^{\Lambda}}{m_2^{\Lambda}} + \frac{m_2^{\Lambda}}{m_1^{\Lambda}} \right)\left( \frac{1}{\xi_{12}}- \xi_{12} \right) \right . \nonumber \\
& & \left . - \frac{m_2^{\Lambda}}{m_3^{\Lambda}}\left( \frac{1}{\xi_{32}}- \xi_{32} \right) - \frac{m_1^{\Lambda}}{m_3^{\Lambda}}\left( \frac{1}{\xi_{13}}- \xi_{13} \right) \right)\cos \delta^{\Lambda} \nonumber \\
& & \left . \left . + \,\frac{1}{4}\left(\frac{1}{\xi_{32}}-\xi_{32} \right)\sin \alpha_2^{\Lambda} \sin\left(\alpha_1^{\Lambda} - \delta^{\Lambda} \right) \right\} \right]\text{cosec}\,\delta^{\Lambda} \sin^2 \theta_{23}^{\Lambda} \sin^2 \theta_{13}^{\Lambda}
\end{eqnarray}
\begin{eqnarray}
\label{dtau}
d_\tau &=& \frac{1}{8}\left[\left( \frac{1}{\xi_{13}}- \xi_{13} \right)\sin\left(\alpha_1^{\Lambda} - \delta^{\Lambda} \right) + \left (\frac{1}{\xi_{32}}-\xi_{32}  \right)\sin\left(\alpha_2^{\Lambda} - \delta^{\Lambda} \right) - \left\{ \frac{m_3^{\Lambda}}{m_2^{\Lambda}}\left( \frac{1}{\xi_{32}}- \xi_{32} \right) \right . \right . \nonumber \\
& & \left . \left . + \,\frac{m_3^{\Lambda}}{m_1^{\Lambda}}\left( \frac{1}{\xi_{13}}- \xi_{13} \right) \right\}\sin\, \delta^{\Lambda} \right]\frac{\sin 2\theta_{12}^{\Lambda} \sin 2\theta_{23}^{\Lambda}} {\sin\theta_{13}^{\Lambda}} \nonumber \\
&+& \frac{1}{2}\left[\left\{\cos^2\theta_{12}^{\Lambda}\left(-\left( \frac{1}{\xi_{32}}- \xi_{32} \right)\cos 2\theta_{23}^{\Lambda}\sin \alpha_2^{\Lambda} - \left( \frac{1}{\xi_{13}}- \xi_{13} \right)\cos^2\theta_{23}^{\Lambda} \times \right . \right . \right . \nonumber \\
& & \sin\left(\alpha_1^{\Lambda} - 2\,\delta^{\Lambda} \right) \bigg) + \left(\left( \frac{1}{\xi_{13}}- \xi_{13} \right)\cos 2\theta_{23}^{\Lambda} \sin \alpha_1^{\Lambda} + \left( \frac{1}{\xi_{32}}- \xi_{32} \right)\cos^2\theta_{23}^{\Lambda} \times \right . \nonumber \\
& & \left . \sin\left(\alpha_2^{\Lambda} - 2\,\delta^{\Lambda} \right) \bigg)\sin^2\theta_{12}^{\Lambda} \bigg\} + \left( \frac{1}{\xi_{12}}- \xi_{12} \right)\sin\left(\alpha_1^{\Lambda} - \alpha_2^{\Lambda} \right) \sin^2\theta_{23}^{\Lambda} \right] \nonumber \\
&+& \left[\frac{1}{4}\left( \frac{1}{\xi_{12}}- \xi_{12} \right)\cos\left(\alpha_1^{\Lambda} - \delta^{\Lambda} \right) \sin \alpha_2^{\Lambda} - \frac{1}{8}\left( \frac{1}{\xi_{12}}- \xi_{12} \right)\left(1 + 2\cos 2\theta_{12}^{\Lambda} \right) \times \right . \nonumber \\
& & \sin\left(\alpha_1^{\Lambda} - \alpha_2^{\Lambda} - \delta^{\Lambda} \right) - \frac{3}{16}\left( \frac{1}{\xi_{32}}- \xi_{32} \right)\sin\left(\alpha_2^{\Lambda} - \delta^{\Lambda} \right) - \frac{1}{4}\left( \frac{1}{\xi_{32}}- \xi_{32} \right) \times \nonumber \\
& & \sin\left(\alpha_2^{\Lambda} - \delta^{\Lambda} \right)\cos 2\theta_{12}^{\Lambda} - \frac{1}{16}\left( \frac{1}{\xi_{32}}- \xi_{32} \right)\sin\left(\alpha_2^{\Lambda} - \delta^{\Lambda} \right)\cos 4\theta_{12}^{\Lambda} - \frac{1}{8}\left( \frac{1}{\xi_{12}}- \xi_{12} \right) \times \nonumber \\
& & \sin\left(\alpha_1^{\Lambda} + \alpha_2^{\Lambda} - \delta^{\Lambda} \right) + \frac{1}{8}\left\{-2\left(\frac{m_1^{\Lambda}} {m_2^{\Lambda}} + \frac{m_2^{\Lambda}} {m_1^{\Lambda}} \right)\left( \frac{1}{\xi_{12}}- \xi_{12} \right) + \frac{3}{2}\frac{m_1^{\Lambda}} {m_3^{\Lambda}}\left( \frac{1}{\xi_{13}}- \xi_{13} \right) \right . \nonumber \\
& & \left . + \,\frac{3}{2}\frac{m_ 2^{\Lambda}}{m_ 3^{\Lambda}}\left( \frac{1}{\xi_ {32}}-\xi_{32} \right) \right\}\sin\, \delta^{\Lambda} + \frac{1}{4}\left\{-\frac{1}{4}\frac{m_1^{\Lambda}}{m_3^{\Lambda}}\left( \frac{1}{\xi_{13}}- \xi_{13} \right) + \frac{1}{4}\frac{m_3^{\Lambda}}{m_2^{\Lambda}}\left( \frac{1}{\xi_{32}}- \xi_{32} \right) \right\} \times \nonumber\\
& & \cos 2 \theta_{12}^{\lambda} \sin\, \delta^{\Lambda} + \frac{1}{16}\left\{\frac{m_3^{\Lambda}}{m_2^{\Lambda}}\left( \frac{1}{\xi_{32}}- \xi_{32} \right) + \frac{1}{4}\frac{m_3^{\Lambda}}{m_1^{\Lambda}}\left( \frac{1}{\xi_{13}}- \xi_{13} \right) \right\}\cos 4\theta_{12}^{\Lambda} \sin\, \delta^{\Lambda}  \nonumber \\
& & + \frac{1}{2}\left( \frac{1}{\xi_{12}}- \xi_{12} \right)\sin\left(\alpha_1^{\Lambda} - \alpha_2^{\Lambda} + \delta^{\Lambda} \right)\sin^2\theta_{12}^{\Lambda} - \frac{1}{2}\left( \frac{1}{\xi_{13}}- \xi_{13} \right)\sin\left(\alpha_1^{\Lambda} - \delta^{\Lambda} \right) \times \nonumber \\
& & \sin^4\theta_{12}^{\Lambda} + \cos 2\theta_{23}^{\Lambda}\left\{\cos^2\theta_{23}^{\Lambda}\left(\frac{1}{2}\left( \frac{1}{\xi_{12}}- \xi_{12} \right)\sin\left(\alpha_1^{\Lambda} - \alpha_2^{\Lambda} + \delta^{\Lambda} \right) + \frac{1}{4}\left( \frac{1}{\xi_{32}}- \xi_{32} \right) \times \right . \right . \nonumber \\
& & \left(-1 + 3\cos 2\theta_{12}^{\Lambda} \right)\sin\left(\alpha_2^{\Lambda} - \delta^{\Lambda} \right) \bigg) + \left(\frac{1}{8}\left(2\left(\frac{m_1^{\Lambda}}{m_2^{\Lambda}} + \frac{m_2^{\Lambda}}{m_1^{\Lambda}} \right)\left( \frac{1}{\xi_{12}}- \xi_{12} \right) - \frac{5}{2}\frac{m_2^{\Lambda}}{m_3^{\Lambda}} \times \right . \right . \nonumber
\end{eqnarray}
\begin{eqnarray}
&& \left . \left( \frac{1}{\xi_{32}}- \xi_{32} \right) - \frac{5}{2}\frac{m_1^{\Lambda}}{m_3^{\Lambda}}\left( \frac{1}{\xi_{13}}- \xi_{13} \right) \right) + \left(\left(\frac{1}{4}\frac{m_1^{\Lambda}}{m_3^{\Lambda}}\left( \frac{1}{\xi_{13}}- \xi_{13} \right) - \frac{1}{4}\frac{m_3^{\Lambda}}{m_2^{\Lambda}}\left( \frac{1}{\xi_{32}}- \xi_{32} \right) \right) \times \right . \nonumber \\
&& \left . \left . \cos 2\theta_{12}^{\Lambda} + \frac{1}{16} \left(\frac{m_3^{\Lambda}}{m_2^{\Lambda}}\left( \frac{1}{\xi_{32}}- \xi_{32} \right) + \frac{m_3^{\Lambda}}{m_1^{\Lambda}}\left( \frac{1}{\xi_{13}}- \xi_{13} \right) \right)\cos 4\theta_{12}^{\Lambda} \right) \right) \sin\, \delta^{\Lambda} \nonumber \\
&& - \,\left(\frac{1}{4}\left( \frac{1}{\xi_{13}}- \xi_{13} \right) \left(1 + 3\cos 2\theta_{12}^{\lambda} \right)\sin\left(\alpha_1^{\Lambda} - \delta^{\Lambda} \right) + \frac{1}{2}\left( \frac{1}{\xi_{12}}- \xi_{12} \right) \times \right . \nonumber \\
&& \left . \left . \sin\left(\alpha_1^{\Lambda} - \alpha_2^{\Lambda} + \delta^{\Lambda} \right) \bigg)\sin^2\theta_{12}^{\Lambda} \right\} \right]\text{cot} \theta_{23}^{\Lambda} \text{cosec}2 \theta_{12}^{\Lambda} \sin\theta_{13}^{\Lambda} \nonumber \\
&+& \left[-\frac{3}{16}\left( \frac{1}{\xi_{13}}- \xi_{13} \right) \cos\left(\alpha_1^{\Lambda} - \delta^{\Lambda} \right) - \frac{5}{16}\left( \frac{1}{\xi_{32}}- \xi_{32} \right) \cos\left(\alpha_2^{\Lambda} - \delta^{\Lambda} \right) + \frac{1}{8}\left\{-\frac{m_1^{\Lambda}}{m_3^{\Lambda}} \times \right . \right . \nonumber \\
&& \left . \left( \frac{1}{\xi_{13}}- \xi_{13} \right) + \left(2\,\frac{m_2^{\Lambda}}{m_1^{\Lambda}}\left( \frac{1}{\xi_{12}}-\xi_{12}\right) - 3\,\frac{m_2^{\Lambda}}{m_3^{\Lambda}}\left( \frac{1}{\xi_{32}}- \xi_{32} \right) \right) \right\}\cos\,\delta^{\Lambda} \nonumber \\
% \end{eqnarray}
% \begin{eqnarray}
&& + \,\frac{1}{4}\left( \frac{1}{\xi_{12}}- \xi_{12} \right) \cos\left(\alpha_1^{\Lambda} - \alpha_2^{\Lambda} + \delta^{\Lambda} \right) - \frac{1}{16}\left( \frac{1}{\xi_{13}}- \xi_{13} \right) \cos\left(\alpha_1^{\Lambda} - 3\,\delta^{\Lambda} \right) \times \nonumber \\
&& \left(-1 + 3\cos 2\theta_{12}^{\Lambda} \right) - \frac{1}{16}\left( \frac{1}{\xi_{32}}- \xi_{32} \right) \cos\left(\alpha_2^{\Lambda} - 3\,\delta^{\Lambda} \right)\left(1 + 3\cos 2\theta_{12}^{\Lambda} \right) + \cos 2\theta_{12}^{\Lambda} \times \nonumber \\
&& \left\{\left(\frac{1}{16}\left( \frac{1}{\xi_{13}}- \xi_{13} \right) + \frac{1}{4}\left( \frac{1}{\xi_{12}}- \xi_{12} \right) \cos \alpha_2^{\Lambda} \right)\cos\left(\alpha_1^{\Lambda} - \delta^{\Lambda} \right) + \frac{1}{16}\left( \frac{1}{\xi_{32}}- \xi_{32} \right) \times \right . \nonumber \\
&& \cos\left(\alpha_2^{\Lambda} - \delta^{\Lambda} \right) + \frac{1}{8}\left(\left(\frac{m_1^{\Lambda}}{m_2^{\Lambda}} + \frac{m_2^{\Lambda}}{m_1^{\Lambda}} \right)\left( \frac{1}{\xi_{12}}- \xi_{12} \right) - \frac{m_2^{\Lambda}}{m_3^{\Lambda}}\left( \frac{1}{\xi_{32}}- \xi_{32} \right) - \frac{m_1^{\Lambda}}{m_3^{\Lambda}} \times \right . \nonumber \\
&& \left . \left . \left . \left( \frac{1}{\xi_{13}}- \xi_{13} \right) \right)\cos\,\delta^{\Lambda} + \frac{1}{4}\left( \frac{1}{\xi_{12}}- \xi_{12} \right)\sin \alpha_2^{\Lambda} \sin\left(\alpha_1^{\Lambda} - \delta^{\Lambda} \right) \right\} \right] \times \nonumber \\
&& \cos^2\theta_{23}^{\Lambda}\text{cosec}\,\delta^{\Lambda}\sin^2\theta_{13}^{\Lambda}
\end{eqnarray}

\begin{equation}
 p_{e1} = d_e + \frac{1}{2} \left[-\left( \frac{1}{\xi_{13}}-\xi_{13} \right)
 \cos^2 \theta^{\Lambda}_{12} \sin(\alpha^{\Lambda}_1 - 2 \delta^{\Lambda}) +
\left( \frac{1}{\xi_{32}}-\xi_{32}\right)\sin(\alpha^{\Lambda}_2 - 2 \delta^{\Lambda}) 
\sin^2 \theta^{\Lambda}_{12} \right]
 \end{equation}
\begin{eqnarray}
 a_{e1} &= & 2 p_{e1}+\left( \frac{1}{\xi_{12}}-\xi_{12} \right) \sin(\alpha^{\Lambda}_1 - \alpha^{\Lambda}_2) 
 \sin^2 \theta^{\Lambda}_{12}  \nonumber \\
&+& \left[ \left( \frac{1}{\xi_{13}}-\xi_{13} \right)
 \sin(\alpha^{\Lambda}_2 - 2 \delta^{\Lambda})-
 \left( \frac{1}{\xi_{12}}-\xi_{12} \right)\sin(\alpha^{\Lambda}_1 - \alpha^{\Lambda}_2) \sin^2 \theta^{\Lambda}_{12}
 \right] \sin^2 \theta^{\Lambda}_{13}
\end{eqnarray}
\begin{eqnarray}
 a_{e2} &=& 2 p_{e1}+ \left( \frac{1}{\xi_{12}}-\xi_{12} \right)  \cos^2 \theta^{\Lambda}_{12}
 \sin(\alpha^{\Lambda}_1 - \alpha^{\Lambda}_2) \nonumber \\
 &-& \left[ 
 \left( \frac{1}{\xi_{12}}-\xi_{12} \right) \cos^2 \theta^{\Lambda}_{12} \sin(\alpha^{\Lambda}_1 - \alpha^{\Lambda}_2)
+ \left( \frac{1}{\xi_{32}}-\xi_{32} \right)\sin(\alpha^{\Lambda}_2 - 2 \delta^{\Lambda}) \right]\sin^2 \theta^{\Lambda}_{13}
\end{eqnarray}

\begin{eqnarray}
 p_{\mu 1} &=& d'_{\mu}- \frac{1}{2} \left[ - \left(\frac{1}{\xi_{13}}-\xi_{13} \right)
 \cos^2 \theta^{\Lambda}_{12}  \sin(\alpha^{\Lambda}_2 - 2 \delta^{\Lambda}) + \left( \frac{1}{\xi_{32}}-\xi_{32} \right)
 \sin(\alpha^{\Lambda}_2 - 2 \delta^{\Lambda}) \sin^2 \theta^{\Lambda}_{12} \right] \sin^2 \theta^{\Lambda}_{23} \nonumber \\
&& + \frac{1}{8} \left[-\left( \frac{1}{\xi_{13}}-\xi_{13} \right) \sin(\alpha^{\Lambda}_1 - \delta^{\Lambda})
 - \left( \frac{1}{\xi_{32}}-\xi_{32} \right)\sin(\alpha^{\Lambda}_2 - \delta^{\Lambda}) \right. \nonumber \\
 &&+ \left. \left[ \frac{m_3}{m_2} \left(\frac{1}{\xi_{32}}-\xi_{32} \right)
  + \frac{m_3}{m_1}\left( \frac{1}{\xi_{13}}-\xi_{13} \right) \right] \sin \delta^{\Lambda}
  \right] \sin 2 \theta^{\Lambda}_{12} \sin 2 \theta^{\Lambda}_{23} \sin \theta^{\Lambda}_{13}
\end{eqnarray}
where $d'_{\mu}$ is the terms appearing in $d_{\mu}$ with coefficent of
$1/\sin \theta^{\Lambda}_{13}$ term set to zero.
\begin{eqnarray}
a_{\mu 1} &=& 2 p_{\mu 1} - \left( \frac{1}{\xi_{12}}-\xi_{12} \right) \cos^2 \theta^{\Lambda}_{23}
 \sin(\alpha^{\Lambda}_1 - \alpha^{\Lambda}_2) \sin^2 \theta^{\Lambda}_{12} \nonumber \\ 
 && -  \frac{1}{2} \Bigg[  \cos \delta^{\Lambda} \Big[ 
 \left( \frac{1}{\xi_{13}}-\xi_{13} \right) \sin \alpha^{\Lambda}_1 + \left( \frac{1}{\xi_{12}}-\xi_{12} \right)\cos 2 \theta^{\Lambda}_{12} 
  \sin(\alpha^{\Lambda}_1 - \alpha^{\Lambda}_2)  \Big] \nonumber \\
  && +\Big[ -\frac{m_1}{m_3}\left( \frac{1}{\xi_{13}}-\xi_{13} \right) 
  +\frac{m_1}{m_2} \left( \frac{1}{\xi_{12}}-\xi_{12} \right) -
  \left( \frac{1}{\xi_{13}}-\xi_{13} \right) \cos \alpha^{\Lambda}_1 - \nonumber 
\end{eqnarray}
\begin{eqnarray} 
&& \left( \frac{1}{\xi_{12}}-\xi_{12} \right) \cos (\alpha^{\Lambda}_1 - \alpha^{\Lambda}_2)
  \Big]\sin \delta^{\Lambda} \Bigg] \sin \theta^{\Lambda}_{13} \sin 2 \theta^{\Lambda}_{23} \tan \theta^{\Lambda}_{12} \\
  && + \Big[ - \left( \frac{1}{\xi_{13}}-\xi_{13} \right) \sin(\alpha^{\Lambda}_1 - 2 \delta^{\Lambda}) 
+ \left( \frac{1}{\xi_{12}}-\xi_{12} \right) \sin(\alpha^{\Lambda}_1 - \alpha^{\Lambda}_2)\sin^2\theta^{\Lambda}_{12}\Big]
\sin^2\theta^{\Lambda}_{13} \sin^2 \theta^{\Lambda}_{23} \nonumber
 \end{eqnarray}
\begin{eqnarray}
 a_{\mu 2} &=& 2 p_{\mu 1} - \left( \frac{1}{\xi_{12}}-\xi_{12} \right) \cos^2 \theta^{\Lambda}_{23}
 \sin(\alpha^{\Lambda}_1 - \alpha^{\Lambda}_2) \cos^2 \theta^{\Lambda}_{12} \nonumber \\ 
 && -  \frac{1}{2} \Bigg[  \cos \delta^{\Lambda} \Big[ 
 \left( \frac{1}{\xi_{32}}-\xi_{32} \right) \sin \alpha^{\Lambda}_2 + \left( \frac{1}{\xi_{12}}-\xi_{12} \right)\cos 2 \theta^{\Lambda}_{12} 
  \sin(\alpha^{\Lambda}_1 - \alpha^{\Lambda}_2)  \Big] \nonumber \\
  && +\Big[ -\frac{m_2}{m_3}\left( \frac{1}{\xi_{32}}-\xi_{32} \right) 
  -\frac{m_2}{m_1} \left( \frac{1}{\xi_{12}}-\xi_{12} \right)+
  \left( \frac{1}{\xi_{32}}-\xi_{32} \right) \cos \alpha^{\Lambda}_2 - \nonumber \\
 && \left( \frac{1}{\xi_{12}}-\xi_{12} \right) \cos (\alpha^{\Lambda}_1 - \alpha^{\Lambda}_2)
  \Big]\sin \delta^{\Lambda} \Bigg] \sin \theta^{\Lambda}_{13} \sin 2 \theta^{\Lambda}_{23} \cot \theta^{\Lambda}_{12} \\
  && + \Big[ \left( \frac{1}{\xi_{32}}-\xi_{32} \right) \sin(\alpha^{\Lambda}_2 - 2 \delta^{\Lambda}) 
+ \left( \frac{1}{\xi_{12}}-\xi_{12} \right) \sin(\alpha^{\Lambda}_1 - \alpha^{\Lambda}_2)\cos^2\theta^{\Lambda}_{12}\Big]
\sin^2\theta^{\Lambda}_{13} \sin^2 \theta^{\Lambda}_{23} \nonumber
\end{eqnarray}
\begin{eqnarray}
 p_{\tau 1} &=& d'_{\tau}- \frac{1}{2} \left[ - \left(\frac{1}{\xi_{13}}-\xi_{13} \right)
 \cos^2 \theta^{\Lambda}_{12}  \sin(\alpha^{\Lambda}_2 - 2 \delta^{\Lambda}) + \left( \frac{1}{\xi_{32}}-\xi_{32}\right)
 \sin(\alpha^{\Lambda}_2 - 2 \delta^{\Lambda}) \sin^2 \theta^{\Lambda}_{12} \right] \cos^2 \theta^{\Lambda}_{23} \nonumber \\
&& - \frac{1}{8} \left[-\left( \frac{1}{\xi_{13}}-\xi_{13} \right) \sin(\alpha^{\Lambda}_1 - \delta^{\Lambda})
 - \left( \frac{1}{\xi_{32}}-\xi_{32} \right)\sin(\alpha^{\Lambda}_2 - \delta^{\Lambda}) \right. \nonumber \\
 &&+ \left. \left[ \frac{m_3}{m_2} \left(\frac{1}{\xi_{32}}-\xi_{32} \right)
  + \frac{m_3}{m_1}\left( \frac{1}{\xi_{13}}-\xi_{13} \right) \right] \sin \delta^{\Lambda}
  \right] \sin 2 \theta^{\Lambda}_{12} \sin 2 \theta^{\Lambda}_{23} \sin \theta^{\Lambda}_{13}
\end{eqnarray}
where $d'_{\tau}$ is the terms appearing in $d_{\tau}$ with coefficient of
$1/\sin \theta^{\Lambda}_{13}$ term set to zero.
\begin{eqnarray}
a_{\tau 1} &=& 2 p_{\tau 1} - \left( \frac{1}{\xi_{12}}-\xi_{12} \right) \sin^2 \theta^{\Lambda}_{23}
 \sin(\alpha^{\Lambda}_1 - \alpha^{\Lambda}_2) \sin^2 \theta^{\Lambda}_{12} \nonumber \\ 
 && +\frac{1}{2} \Bigg[  \cos \delta^{\Lambda} \Big[ 
 \left( \frac{1}{\xi_{13}}-\xi_{13} \right) \sin \alpha^{\Lambda}_1 + 
 \left( \frac{1}{\xi_{12}}-\xi_{12} \right)\cos 2 \theta^{\Lambda}_{12} 
  \sin(\alpha^{\Lambda}_1 - \alpha^{\Lambda}_2)  \Big] \nonumber \\
  && +\Big[ -\frac{m_1}{m_3}\left( \frac{1}{\xi_{13}}-\xi_{13} \right) 
  +\frac{m_1}{m_2} \left( \frac{1}{\xi_{12}}-\xi_{12} \right) -
  \left( \frac{1}{\xi_{13}}-\xi_{13} \right) \cos \alpha^{\Lambda}_1 - \nonumber \\
 && \left( \frac{1}{\xi_{12}}-\xi_{12} \right) \cos (\alpha^{\Lambda}_1 - \alpha^{\Lambda}_2)
  \Big]\sin \delta^{\Lambda} \Bigg] \sin \theta^{\Lambda}_{13} \sin 2 \theta^{\Lambda}_{23} \tan \theta^{\Lambda}_{12}\\
  && + \Big[ - \left( \frac{1}{\xi_{13}}-\xi_{13} \right) \sin(\alpha^{\Lambda}_1 - 2 \delta^{\Lambda}) 
+ \left( \frac{1}{\xi_{12}}-\xi_{12} \right) \sin(\alpha^{\Lambda}_1 - \alpha^{\Lambda}_2)\sin^2\theta^{\Lambda}_{12}\Big]
\sin^2\theta^{\Lambda}_{13} \cos^2 \theta^{\Lambda}_{23} \nonumber 
 \end{eqnarray}
 \begin{eqnarray}
 a_{\tau 2} &=& 2 p_{\tau 1} - \left( \frac{1}{\xi_{12}}-\xi_{12} \right) \sin^2 \theta^{\Lambda}_{23}
 \sin(\alpha^{\Lambda}_1 - \alpha^{\Lambda}_2) \cos^2 \theta^{\Lambda}_{12} \nonumber \\ 
 && -  \frac{1}{2} \Bigg[  \cos \delta^{\Lambda} \Big[ 
 -\left( \frac{1}{\xi_{32}}-\xi_{32} \right) \sin \alpha^{\Lambda}_2 - \left( \frac{1}{\xi_{12}}-\xi_{12} \right)\cos 2 \theta^{\Lambda}_{12} 
  \sin(\alpha^{\Lambda}_1 - \alpha^{\Lambda}_2)  \Big] \nonumber \\
  && +\Big[ \frac{m_2}{m_3}\left( \frac{1}{\xi_{32}}-\xi_{32} \right) 
  -\frac{m_2}{m_1} \left( \frac{1}{\xi_{12}}-\xi_{12} \right)+
  \left( \frac{1}{\xi_{32}}-\xi_{32} \right) \cos \alpha^{\Lambda}_2 +\nonumber \\
 && \left( \frac{1}{\xi_{12}}-\xi_{12} \right) \cos (\alpha^{\Lambda}_1 - \alpha^{\Lambda}_2)
  \Big]\sin \delta^{\Lambda} \Bigg] \sin \theta^{\Lambda}_{13} \sin 2 \theta^{\Lambda}_{23} \cot \theta^{\Lambda}_{12}  \\
  && - \Big[ \left( \frac{1}{\xi_{32}}-\xi_{32} \right) \sin(\alpha^{\Lambda}_2 - 2 \delta^{\Lambda}) 
- \left( \frac{1}{\xi_{12}}-\xi_{12} \right) \sin(\alpha^{\Lambda}_1 - \alpha^{\Lambda}_2)\cos^2\theta^{\Lambda}_{12}\Big]
\sin^2\theta^{\Lambda}_{13} \cos^2 \theta^{\Lambda}_{23} \nonumber
\end{eqnarray}
\begin{eqnarray}
  p_{e 2} &=& \Bigg[ \frac{1}{4} \left( \frac{1}{\xi_{13}}-\xi_{13} \right)
 \sin(\alpha^{\Lambda}_1 - \delta^{\Lambda}) 
 + \frac{1}{4} \left( \frac{1}{\xi_{32}}-\xi_{32} \right)
  \sin(\alpha^{\Lambda}_2 - \delta^{\Lambda}) \nonumber \\
 && + \Big[ \frac{1}{4} \frac{m_3}{m_2} 
 \left( \frac{1}{\xi_{32}}-\xi_{32} \right) + \frac{1}{4} \frac{m_3}{m_1} 
 \left( \frac{1}{\xi_{13}}-\xi_{13} \right) \Big] \sin \delta^{\Lambda}
  \Bigg] \sin \theta^{\Lambda}_{13} \sin 2 \theta^{\Lambda}_{12} \cot \theta^{\Lambda}_{23}  \\
&& - 2 \Bigg[-\frac{1}{4} \left( \frac{1}{\xi_{13}}-\xi_{13} \right) 
  \cos^2 \theta^{\Lambda}_{12} \sin(\alpha^{\Lambda}_1 -2 \delta^{\Lambda})+
  \frac{1}{4} \left( \frac{1}{\xi_{32}}-\xi_{32} \right) 
  \sin^2 \theta^{\Lambda}_{12} \sin(\alpha^{\Lambda}_2 -2 \delta^{\Lambda})+ \Bigg]\sin^2\theta^{\Lambda}_{13} \nonumber
\end{eqnarray}
\begin{eqnarray}
  p_{e 3} &=&-\Bigg[ \frac{1}{4} \left( \frac{1}{\xi_{13}}-\xi_{13} \right)
 \sin(\alpha^{\Lambda}_1 - \delta^{\Lambda}) 
 + \frac{1}{4} \left( \frac{1}{\xi_{32}}-\xi_{32} \right)
  \sin(\alpha^{\Lambda}_2 - \delta^{\Lambda}) \nonumber \\
 && + \Big[ \frac{1}{4} \frac{m_3}{m_2} 
 \left( \frac{1}{\xi_{32}}-\xi_{32} \right) + \frac{1}{4} \frac{m_3}{m_1} 
 \left( \frac{1}{\xi_{13}}-\xi_{13} \right) \Big] \sin \delta^{\Lambda}
  \Bigg] \sin \theta^{\Lambda}_{13} \sin 2 \theta^{\Lambda}_{12} \tan \theta^{\Lambda}_{23}  \\
&& - 2 \Bigg[-\frac{1}{4} \left( \frac{1}{\xi_{13}}-\xi_{13} \right) 
  \cos^2 \theta^{\Lambda}_{12} \sin(\alpha^{\Lambda}_1 -2 \delta^{\Lambda})+
  \frac{1}{4} \left( \frac{1}{\xi_{32}}-\xi_{32} \right) 
  \sin^2 \theta^{\Lambda}_{12} \sin(\alpha^{\Lambda}_2 -2 \delta^{\Lambda})\Bigg]\sin^2\theta^{\Lambda}_{13} \nonumber
\end{eqnarray}
\begin{eqnarray}
  p_{\mu 2} &=& \frac{1}{2} \cos^2 \theta^{\Lambda}_{23} \Bigg[ 
 \left( \frac{1}{\xi_{32}}-\xi_{32} \right) \cos^2 \theta^{\Lambda}_{12}  \sin \alpha^{\Lambda}_2- 
\left( \frac{1}{\xi_{13}}-\xi_{13} \right) \sin \alpha^{\Lambda}_1 \sin^2 \theta^{\Lambda}_{12}  \Bigg]\nonumber \\
&& + \frac{1}{4} \Bigg[ - \left( \frac{1}{\xi_{13}}-\xi_{13} \right) \sin(\alpha^{\Lambda}_1 - \delta^{\Lambda})
- \left( \frac{1}{\xi_{32}}-\xi_{32} \right) \sin(\alpha^{\Lambda}_2 - \delta^{\Lambda})\Bigg] \sin 2 \theta^{\Lambda}_{12} 
\sin 2 \theta^{\Lambda}_{23} \sin \theta^{\Lambda}_{13}  \nonumber \\
&& + \frac{1}{2} \Bigg[  - \left( \frac{1}{\xi_{13}}-\xi_{13} \right) \cos^2 \theta^{\Lambda}_{12}
\sin(\alpha^{\Lambda}_1 - \delta^{\Lambda}) + \nonumber \\
& & \left( \frac{1}{\xi_{32}}-\xi_{32} \right) \sin(\alpha^{\Lambda}_2 - \delta^{\Lambda})
\sin^2 \theta^{\Lambda}_{12} \Bigg]\sin^2 \theta^{\Lambda}_{23} \sin^2 \theta^{\Lambda}_{13} 
\end{eqnarray}
\begin{eqnarray}
 p_{\tau 2} &=& -  \frac{1}{4} \Bigg[ 
 \left( \frac{1}{\xi_{32}}-\xi_{32} \right) \cos^2 \theta^{\Lambda}_{12}  \sin \alpha^{\Lambda}_2- 
\left( \frac{1}{\xi_{13}}-\xi_{13} \right) \sin \alpha^{\Lambda}_1 \sin^2 \theta^{\Lambda}_{12} 
 \Bigg] \sin 2 \theta^{\Lambda}_{23} \cot \theta^{\Lambda}_{23} \nonumber \\
 && + \frac{1}{4} \Bigg[ \Big[ -
 \left( \frac{1}{\xi_{13}}-\xi_{13} \right) \sin(\alpha^{\Lambda}_1 - \delta^{\Lambda})
 - \left( \frac{1}{\xi_{32}}-\xi_{32} \right) \sin(\alpha^{\Lambda}_2 - \delta^{\Lambda})\Big] \cos 2 \theta^{\Lambda}_{23}\nonumber \\
 && - \Big[  \frac{m_3}{m_2} 
 \left( \frac{1}{\xi_{32}}-\xi_{32} \right) + \frac{m_3}{m_1} 
 \left( \frac{1}{\xi_{13}}-\xi_{13} \right)\Big] \sin \delta^{\Lambda}
 \Bigg] \sin 2 \theta^{\Lambda}_{12} \cot \theta^{\Lambda}_{23} \sin \theta^{\Lambda}_{13}  \\
 && - \frac{1}{4} \Bigg[ \left( \frac{1}{\xi_{13}}-\xi_{13} \right) 
  \cos^2 \theta^{\Lambda}_{12} \sin(\alpha^{\Lambda}_1 -2 \delta^{\Lambda})+ \nonumber \\
& &\left( \frac{1}{\xi_{32}}-\xi_{32} \right) 
  \sin^2 \theta^{\Lambda}_{12} \sin(\alpha^{\Lambda}_2 -2 \delta^{\Lambda})
 \Bigg] \cot \theta^{\Lambda}_{23} \sin 2 \theta^{\Lambda}_{23} \sin^2 \theta^{\Lambda}_{13} \nonumber
\end{eqnarray}
\begin{eqnarray}
  p_{\tau 3} &=& \frac{1}{2} \Bigg[  
  \left( \frac{1}{\xi_{32}}-\xi_{32} \right) \cos^2 \theta^{\Lambda}_{12}  \sin \alpha^{\Lambda}_2- 
\left( \frac{1}{\xi_{13}}-\xi_{13} \right) \sin \alpha^{\Lambda}_1 
\sin^2 \theta^{\Lambda}_{12} \Bigg] \sin^2 \theta^{\Lambda}_{23} \nonumber \\
&& +  \frac{1}{4} \Bigg[  \left( \frac{1}{\xi_{13}}-\xi_{13} \right) \sin(\alpha^{\Lambda}_1 - \delta^{\Lambda})
+ \left( \frac{1}{\xi_{32}}-\xi_{32} \right) \sin(\alpha^{\Lambda}_2 - \delta^{\Lambda})
\Bigg] \sin 2 \theta^{\Lambda}_{12} 
\sin 2 \theta^{\Lambda}_{23} \sin \theta^{\Lambda}_{13}  \nonumber \\
&& \frac{1}{2} \Bigg[
\Big[ \left( \frac{1}{\xi_{13}}-\xi_{13} \right) \cos \alpha^{\Lambda}_1 \cos^2 \theta^{\Lambda}_{12}
- \left( \frac{1}{\xi_{32}}-\xi_{32} \right) \cos \alpha^{\Lambda}_2 \sin^2 \theta^{\Lambda}_{12}
\Big] \sin 2 \delta^{\Lambda} \\
&& - \Big[ \left( \frac{1}{\xi_{13}}-\xi_{13} \right) \sin \alpha^{\Lambda}_1 \cos^2 \theta^{\Lambda}_{12}
- \left( \frac{1}{\xi_{32}}-\xi_{32} \right) \sin \alpha^{\Lambda}_2 \sin^2 \theta^{\Lambda}_{12}
\Big] \cos 2 \delta^{\Lambda}
\Bigg]\cos^2 \theta^{\Lambda}_{23} \sin^2 \theta^{\Lambda}_{13} \nonumber 
\end{eqnarray}

% \bibliographystyle{apsrev}
% \bibliography{ref,neutrino}
% %

\end{document}